\documentclass[aps,nofootinbib,noshowpacs,showkeys,preprintnumbers,amsmath,amssymb]{revtex4}
\usepackage{graphicx,color}
\usepackage{graphics}
\usepackage{rotating}
\usepackage{amssymb}

\usepackage{color}

\usepackage{epsfig}
\usepackage{dcolumn}
\usepackage{bm}
\usepackage{multirow}

\def\be{\begin{equation}}
\def\ee{\end{equation}}
\def\ba{\begin{eqnarray}}
\def\ea{\end{eqnarray}}
\def\bea{\begin{eqnarray}}
\def\eea{\end{eqnarray}}
\def\bes{\begin{subequations}}
\def\ees{\end{subequations}}
\def\bear{\begin{array}}
\def\eear{\end{array}}

\newcommand{\A}{{\mathcal{A}}}
\newcommand{\tA}{{\widetilde {\mathcal{A}}}}
\newcommand{\ta}{{\widetilde a}}
\newcommand{\tal}{{\widetilde \alpha}}

\newcommand{\tK}{{\widetilde K}}
\newcommand{\td}{{\widetilde d}}

\newcommand{\MSbar}{\overline{\rm MS}}

\begin{document}

\title{Renormalon-based resummation of Bjorken polarised sum rule in holomorphic QCD}
\author{C\'esar Ayala$^a$}
\email{c.ayala86@gmail.com}
\author{Camilo Castro-Arriaza$^b$}
\email{camilo.castroa@sansano.usm.cl}
\author{Gorazd Cveti\v{c}$^b$}
\email{gorazd.cvetic@gmail.com}
\affiliation{$^a$Instituto de Alta Investigaci\'on, Sede La Tirana, Universidad de Tarapac\'a, Av.~La Tirana 4802, Iquique, Chile}
\affiliation{$^b$Department of Physics, Universidad T{\'e}cnica Federico Santa Mar{\'\i}a, Avenida España 1680, Valpara{\'\i}so, Chile}

\date{\today}

\begin{abstract}
Approximate knowledge of the renormalon structure of the Bjorken polarised sum rule (BSR) ${\overline \Gamma}_1^{{\rm p-n}}(Q^2)$ leads to the corresponding BSR characteristic function that allows us to evaluate the leading-twist part of BSR. In our previous work \cite{pPLB}, this evaluation (resummation) was performed using perturbative QCD (pQCD) coupling $a(Q^2) \equiv \alpha_s(Q^2)/\pi$ in specific renormalisation schemes. In the present paper, we continue this work, by using instead holomorphic couplings [$a(Q^2) \mapsto {\mathcal A}(Q^2)$] that have no Landau singularities and thus require, in contrast to the pQCD case,  no regularisation of the resummation formula. The $D=2$ and $D=4$ terms are included in the Operator Product Expansion (OPE) of inelastic BSR, and fits are performed to the available experimental data in a specific interval $(Q^2_{\rm min},Q^2_{\rm max})$ where $ Q^2_{\rm max}=4.74 \ {\rm GeV}^2$. We needed relatively high $Q^2_{\rm min} \approx 1.7 \ {\rm GeV}^2$ in the pQCD case since the pQCD coupling $a(Q^2)$ has Landau singularities at $Q^2 \lesssim 1 \ {\rm GeV}^2$. Now, when holomorphic (AQCD) couplings ${\mathcal A}(Q^2)$ are used, no such problems occur: for the $3 \delta$AQCD and $2 \delta$AQCD variants the preferred values are $Q^2_{\rm min} \approx 0.6 \ {\rm GeV}^2$. The preferred values of $\alpha_s$ in general cannot be unambiguously extracted, due to large uncertainties of the experimental BSR data. At a fixed value of $\alpha_s^{{\overline {\rm MS}}}(M_Z^2)$, the values of the $D=2$ and $D=4$ residue parameters are determined in all cases, with the corresponding uncertainties.
\end{abstract}
\maketitle

\section{Introduction}
\label{sec:intr}

We analyse in this work the inelastic polarised Bjorken sum rule (BSR)  ${\overline {\Gamma}}_1^{{\rm p-n}}(Q^2)$ \cite{BjorkenSR, BSR70}, which is the difference of the $g_1$ spin-dependent structure functions of the proton and neutron integrated over the $x$-Bjorken parameter range $0 < x < 1$. Its Operator Product Expansion (OPE) has a relatively simple form, because it is an isovector and spacelike quantity.

Experimental results for ${\overline {\Gamma}}_1^{{\rm p-n}}(Q^2)$, though often with significant statistical and systematic uncertainties, are available from various experiments: CERN \cite{CERN}, DESY \cite{DESY}, SLAC \cite{SLAC}, and Jefferson Lab \cite{JeffL1,JeffL2,JeffL3,JeffL4,JeffL5}. These experimental results are based on the measured values of the spin-dependent structure functions over various values of Bjorken $x$, and over various values of $Q^2$ in the wide range $0.02 \ {\rm GeV}^2 < Q^2 < 5 \ {\rm GeV}^2$ where $q^2 \equiv - Q^2$ is the squared momentum transfer.

On the other hand, the considered inelastic BSR ${\overline \Gamma}_1^{{\rm p-n}}(Q^2)$ is evaluated theoretically by using OPE that is usually truncated at the dimension $D=2$ ($\sim 1/Q^2$) or $D=4$ ($\sim 1/(Q^2)^2$) term. The leading-twist ($D=0$) term has the canonical QCD part $d(Q^2)$ which is in general evaluated as a truncated perturbation series (TPS) in powers of the pQCD coupling $a(Q^2) = \alpha_s(Q^2)/\pi$. The values of the first four expansion coefficients (i.e., up to $\sim a^4$) have been calculated exactly \cite{GorLar1986,LarVer1991,BaiCheKu2010}, and the value of the coefficient at $a^5$ can be estimated \cite{KatStar}. In this way, the mentioned OPE expression is then fitted to the experimental data, and the values of the OPE coefficient of the $D=2$ term, and possibly of the $D=4$ term, can be extracted. This pQCD OPE approach, or specific variants of it, has been followed in the works \cite{ABFR,JeffL1,JeffL2,JeffL4,ACKS,KoMikh,LFHBSR,BSRPMC}. 

The described approaches of theoretical evaluation do not involve direct resummations in the canonical QCD part of BSR, $d(Q^2).$\footnote{In \cite{KoMikh,LFHBSR,BSRPMC} the resummations are indirect, by fixing (various) renormalisation scale(s) in the TPS according to different criteria.} Analyses of BSR by some of us were also performed in the approaches that use holomorphic QCD (AQCD) running couplings $a(Q^2) \mapsto \A(Q^2)$ (i.e., free of Landau singularities) in $d(Q^2)$, enabling the evaluation of $d(Q^2)$ even at lower $Q^2 \lesssim 1 \ {\rm GeV}^2$ \cite{ACKS}. However, even in those AQCD cases, the series for $d(Q^2)$ were truncated. Actually, we know that $d(Q^2)$ of BSR has the expansion coefficients $d_n$ (at $a^{n+1}$) that grow very fast with increasing $n$, approximately as $d_n \sim n! (\beta_0/p)^n$ with $p=1$ [due to the leading infrared renormalon close to the origin, i.e., the pole of the Borel transform ${\cal B}_d(u)$ at $u=1$], which indicates that truncation of $d(Q^2)$, at $\sim a^4$ or $\sim a^5$, may miss important contributions.

In our previous work \cite{pPLB}, we evaluated the QCD canonical part $d(Q^2)$ of BSR by performing a renormalon-based resummation. The extension of the $d(Q^2)$ expansion to higher powers of $a$ was performed by a renormalon-based approach. The latter approach allows us to construct a characteristic function of $d(Q^2)$ and enables us to perform resummation in a form of integration that involves $a(Q^{'2} = t Q^{2})$ over the entire range of spacelike scales ($0 < t < \infty$), cf.~Eq.~(\ref{dres2b}). In \cite{pPLB} the pQCD coupling $a(Q^{'2})$ was used in the resummation formula and fits of the corresponding (truncated) OPE to the experimental data were performed. The pQCD coupling has Landau singularities (in the range $0 < Q^{'2} < 1 \ {\rm GeV}^2$), and hence a regularisation of the resummed result was needed. In the present work, which can be regarded as a continuation of our previous work \cite{pPLB}, we perform the resummation of $d(Q^2)$ instead by using holomorphic (analytic, AQCD) couplings $\A(Q^{'2})$. These running couplings have no Landau singularities, and thus no regularisation is needed. Subsequently, we perform fits with the corresponding (truncated) OPE for BSR ${\overline {\Gamma}}_1^{{\rm p-n}}(Q^2)$ with the experimental data.

We point out that other authors have worked on applications of infrared-finite (Landau-singularities-free) QCD couplings, primarily applied to (inelastic) BSR. For example, inelastic BSR ${\overline \Gamma}_1^{{\rm p-n}}(Q^2)$ at very low $Q^2$ ($0 \leq Q^2 \lesssim 1 \ {\rm GeV}^2$) was evaluated by expansions \cite{JeffL2} motivated by chiral perturbation theory or by the light-front holographic QCD (LFH) effective coupling \cite{LFH1,LFHBSR}. BSR in a wider range $0 \leq Q^2 \lesssim 10 \ {\rm GeV}^2$ was evaluated with an effective coupling from Schwinger-Dyson (SD) approach in \cite{DSBSR} and with an extended effective charge LFH coupling in \cite{LFHBSRext}. These latter two effective couplings (SD, LFH) were considered to be de facto effective charges for the (inelastic) BSR, so that no higher-twist OPE terms were included. While these couplings are also IR-finite as our AQCD holomorphic couplings, and thus contain nonperturbative physics, there are several differences between them and our approach. In our approach, we are not in (de facto) effective charge scheme (ECH) \cite{Grunberg} of BSR, but in various perturbatively well defined $\MSbar$-type renormalisation schemes (such as P44-schemes, see later in the text). Therefore, the corresponding beta-function coefficients $\beta_n$ are not increasing singularly (factorially) with $n$ and thus do not eliminate the renormalon behaviour of the BSR perturbation coefficients,\footnote{The renormalon growth of the coefficients $d_n$ can be eliminated not just by a special choice of a renormalisation scheme such as ECH, but also, e.g., by a specific scale setting at each order of expansion (PMC: Principle of Maximal Conformality) \cite{PMC}, see also \cite{LFHBSR,BSRPMC}.}
$d_n \sim n! \beta_0^n$.  In that sense, our approach is somewhat more conservative, and is tied to the perturbative theoretical knowledge and renormalon structure of $d(Q^2)$ of BSR. Due to this behaviour, we need to perform a resummation of terms $d_n \A_n(Q^2)$ [these are AQCD analogues of the pQCD terms  $d_n a(Q^2)^n$] over all $n$ in $d(Q^2)$. This implies that we also need to include higher-twist terms ($D > 0$) in the OPE because, in general, we cannot expect that all the nonperturbative BSR effects are contained in (the resummation of) the terms  $d_n \A_n(Q^2)$. As a consequence, the condensate parameters of the higher-twist terms, whose values are to be extracted from the fit procedure, become significantly dependent on how the evaluation (resummation) of the leading-twist contribution $d(Q^2)$ is made. For example, in pQCD there is an intrinsic renormalon ambiguity in the evaluation of $d(Q^2)$ that is then reflected in the higher-twist parameter value ambiguity. In our resummation in AQCD, no such renormalon ambiguity appears; nonetheless, the parametrisation of the AQCD couplings $\A(Q^2)$ in the IR-regime is ambiguous\footnote{For example, the variation of our threshold scale $M_1$ of the spectral function of $\A(Q^2)$ is a measure of such ambiguity, cf.~Eq.~(\ref{rhoAnd}).}, and it is this ambiguity that is then reflected as an important (and unremovable) theoretical uncertainty in the extracted values of the higher-twist parameters. Yet another difference between our (resummed) AQCD approach and the aforementioned effective charge approach (LFH, SD) is that we do not try to describe with our approach the values of BSR all the way down to $Q^2 \to 0$, but rather to describe moderately low ranges $0.5 \ {\rm GeV}^2 < Q^2$ ($< {\cal O}(10 \ {\rm GeV}^2)$), i.e., the ranges which are inaccessible with approaches that use the pQCD coupling $a(Q^2)$. Besides extending the applicability range of $Q^2$, another aim of this work is to check whether the inclusion of some of the nonperturbative effects in the leading-twist term $d(Q^2)$ reduces the experimental uncertainties of the extracted higher-twist parameters as compared to the case when pQCD coupling is used.

In Sec.~\ref{sec:theor} we summarise some of the theoretical aspects of the approach, though we refer for details to our previous works \cite{renmod,pPLB}. In Sec.~\ref{sec:AQCDcan} we present the numerical results of the various evaluations of the canonical part of BSR, $d(Q^2)$, for the two types of the holomorphic (AQCD) coupling applied. In particular, we explore the numerical behaviour of $d(Q^2)$ when it is either resummed, or has a form of approximants based on truncated information, for these AQCD couplings. We then use the resummed AQCD results for $d(Q^2)$ in Sec.~\ref{sec:fit} to evaluate the (truncated) OPE of BSR ${\overline {\Gamma}}_1^{{\rm p-n}}(Q^2)$ and fit it to the experimental data, and thus we extract the values of the OPE $D=2$ and $D=4$ coefficients, for the two types of AQCD coupling used. In Sec.~\ref{sec:fit} we also compare the obtained results with those of Ref.~\cite{pPLB} where pQCD coupling is used. Finally, in Sec.~\ref{sec:concl} we discuss the obtained numerical results, and summarise the work. Some formal details used in this work are given in Appendices \ref{app:AQCD}-\ref{app:expu}, and an extension of our fit approach is discussed in Appendix \ref{app:ks}.

\section{Theoretical expressions}
\label{sec:theor}

In this Section we only summarise briefly the final results of the formalism explained in general in Ref.~\cite{renmod} and, when applied to BSR in particular, in Ref.~\cite{pPLB}.

The theoretical OPE expression for the inelastic BSR ${\overline \Gamma}_1^{{\rm p-n}}(Q^2)$, truncated at the dimension $D=4$ term ($\sim 1/(Q^2)^2$), has the form
\bea
\label{BSROPE}
 {\overline \Gamma}_1^{{\rm p-n}, \rm {OPE}}(Q^2) &=& {\Big |}\frac{g_A}{g_V} {\Big |} \frac{1}{6} \left[ 1 - d(Q^2) - \delta d(Q^2)_{m_c} \right]
\nonumber\\
&+& \frac{M_N^2}{9} \frac{\left[ (a_2^{({\rm p-n})}(Q^2) + 4 d_2^{({\rm p-n})}(Q^2))+ 4 {\bar f}_2 \; {a}(Q^2)^{{k_1}} \right]}{Q^2} + \frac{\mu_6}{(Q^2)^2} .
\eea
Here $|g_A/g_V|$ is the ratio of the nucleon axial charge, and we take the value $|g_A/g_V|=1.2754$ \cite{PDG2023}. In the $D=2$ term,  $k_1=32/81$  \cite{Kawetal1996} is the anomalous dimension of the $D=2$ operator; $M_N =0.9389$ GeV is the nucleon mass, and $(a_2^{{\rm p-n}} + 4 d_2^{{\rm p-n}})$ is a combination of the twist-2 target correction and of a twist-3 matrix element.\footnote{For an earlier estimate of the $\sim 1/Q^2$ contribution to BSR, see \cite{NSV}.}
In pQCD, its one-loop running is known \cite{Bali1,Bali2}
\be
(a_2^{({\rm p-n})}(Q^2) + 4 d_2^{({\rm p-n})}(Q^2)) =
a_2^{({\rm p-n})}(Q_0^2) \left(\frac{a(Q^2)}{a(Q_0^2)}\right)^{B_{a2}} + 
  d_2^{({\rm p-n})}(Q_0^2) \left(\frac{a(Q^2)}{a(Q_0^2)}\right)^{B_{d2}},
 \label{a2d2p} \ee
 where the corresponding anomalous dimensions (for $N_f=3$) are\footnote{We thank G.~Bali for clarification on $B_{a2}$.} $B_{a2}=50/81$ \cite{Bali1} and $B_{d2}=77/81$ \cite{Bali2}. We note that $k_1$, $B_{a2}$ and $B_{d2}$ are one-loop anomalous dimensions and therefore are renormalisation scheme independent. The values at $Q_0^2=4 \ {\rm GeV}^2$ were obtained in \cite{Bali2}\footnote{In our convention, $a_2$ and $d_2$ are the authors' $a_2/2$ and $d_2/2$.}: $a_2(Q_0^2)=0.0311 \pm 0.0095$ and $4 d_2(Q_0^2)=0.0228 \pm 0.0196$. The analogue of the pQCD power $a(Q^2)^{\nu}$ in AQCD is $\A_{\nu}(Q^2)$. This, in the leading order, is equal to $\tA_{\nu}(Q^2)$ ($\A_{\nu} = \tA_{\nu} + {\cal O}(\tA_{\nu+1})$), where $\tA_{\nu}(Q^2)$ is the extension of the logarithmic derivative Eq.~(\ref{tAn}) to noninteger values of $\nu$. The quantities $\tA_{\nu}$ and $\A_{\nu}$, in general in AQCD, were constructed in \cite{GCAK}. All this then implies that in the AQCD the running $(a_2+4 d_2)$ quantity has the form
\be
(a_2^{({\rm p-n})}(Q^2) + 4 d_2^{({\rm p-n})}(Q^2)) =
(0.0311 \pm 0.0095) \frac{\tA_{B_{a2}}(Q^2)}{\tA_{B_{a2}}(Q_0^2)} + 
(0.0228 \pm 0.0196) \frac{\tA_{B_{d2}}(Q^2)}{\tA_{B_{d2}}(Q_0^2)},
\label{a2d2A} \ee
where $Q_0^2 = 4 \ {\rm GeV}^2$. Similar kinds of quantities in AQCD were encountered in \cite{CAGCLG} in the context of QCD effects in the $0\nu\beta\beta$-decay. Similarly, the pQCD expression ${a}(Q^2)^{{k_1}}$ in the $D=2$ term in OPE (\ref{BSROPE}) is in AQCD replaced by $\tA_{k_1}(Q^2)$ ($k_1=32/81$). 

The quantity $d(Q^2)$ is the canonical $N_f=3$ massless QCD contribution whose power expansion in terms of the (pQCD) coupling $a(Q^2) \equiv \alpha_s(Q^2)/\pi$ is
\bea
d(Q^2)_{\rm pt} & = & a(Q^2) + d_1 a(Q^2)^2 + d_2 a(Q^2)^3 + d_3 a(Q^2)^4 + {\cal O}(a^5),
\label{dptkap1}
\eea
The first coefficients $d_j$ ($j=0,1,2,3$) are exactly known and were obtained  in the $\MSbar$ scheme  in \cite{GorLar1986,LarVer1991,BaiCheKu2010}, while the next coefficient $d_4$ can be estimated by the effective charge (ECH) method \cite{KatStar}, and its value in the 5-loop $\MSbar$ scheme is $d_4^{{\MSbar}}=1557.4$. We will take $d_4^{\MSbar}\approx 1557.4 \pm 32.8$, cf.~\cite{pPLB}.

The term $\delta d(Q^2)_{m_c}$ in Eq.~(\ref{BSROPE}) is the correction due to the non-decoupling of the charm mass (i.e., $m_c \not= \infty$ effects) \cite{Blumetal}, it is $\sim a(Q^2)^2$ and is written down in \cite{pPLB}, and the term $a(Q^2)^2$ is now in AQCD replaced by $\tA_2(Q^2)$.

 The value of the dimensionless parameter ${\bar f}_2$, and possibly of the coefficient $\mu_6$ at the $D=4$ OPE term, are to be determined by the fitting of the theoretical expression (\ref{BSROPE}) to the BSR experimental values. Due to the lack of theoretical knowledge, $\mu_6$ will be considered to be $Q^2$-independent. The OPE (\ref{BSROPE}) will be truncated either at $D=2$ or $D=4$ term.

As explained in Refs.~\cite{renmod,pPLB}, the construction of the renormalon-motivated resummation is largely based on the idea of reorganising the perturbation expansion (\ref{dptkap1}) of $d(Q^2)$ in powers of $a(\mu^2)$ into an expansion in logarithmic derivatives ${\ta}_{n+1}$
\be
{\ta}_{n+1}(\mu^2) \equiv \frac{(-1)^n}{n! \beta_0^n} \left( \frac{d}{d \ln \mu^2} \right)^n {a}(\mu^2) \qquad (n=0,1,2,\ldots),
\label{tan} \ee
leading to
\bes
\label{dptdlpt}
\bea
d(Q^2) &=& a(\kappa Q^2) + d_1(\kappa) a(\kappa Q^2)^2 + \ldots + d_n(\kappa) a(\kappa Q^2)^{n+1} + \ldots
\label{dpt} \\
& = & a( \kappa Q^2) + {\td}_1(\kappa) \; {{\ta}}_2(\kappa Q^2) + \ldots + {\td}_n(\kappa) \; {{\ta}}_{n+1}(\kappa Q^2) + \ldots.
\label{dlpt} \eea \ees
Here $\kappa \equiv \mu^2/Q^2$ ($0 < \kappa \lesssim 1$), where $\mu^2$ is a chosen renormalisation scale. This reorganisation gives us new coefficients $\td_n(\kappa)$ that are linear combinations of the original expansion coefficients $d_n(\kappa), \ldots, d_1(\kappa)$. It is these new coefficients $\td_n(\kappa)$ that play the central role in the construction of the renormalon-motivated resummation. This approach can be also interpreted as an extension of the large-$\beta_0$ approach of resummation of Neubert \cite{Neubert} to all loops. For the large-$\beta_0$ structure of the BSR, see \cite{BK1993,Renormalons} (cf.~also \cite{Kat1,Kat2}).

The renormalon-resummed value of the canonical part $d(Q^2)$ is
 \be
d(Q^2)_{\rm res} = {\rm Re} \left[
\int_0^\infty \frac{dt}{t} G_d(t) {a}(t e^{-\tK} Q^2+ i \varepsilon) \right]
\label{dres2b}
\ee
when using the pQCD running coupling $a(\mu^2)$, and
\be
d(Q^2)_{\rm res, AQCD} = \int_0^\infty \frac{dt}{t} G_d(t) {\A}(t e^{-\tK} Q^2) .
\label{dres2c}
\ee
when using the holomorphic (AQCD) IR-safe running coupling ${a}(Q^2) \mapsto {\A}(Q^2)$, i.e., a coupling that has no Landau singularities but practically coincides with ${a}(Q^2)$ at large $|Q^2| > \Lambda^2_{\rm QCD}$. For example, such are the 2$\delta$AQCD \cite{2dAQCD,2dAQCDb} or 3$\delta$AQCD couplings \cite{3dAQCD,3dAQCDb}. In such a case, no regularisation is needed in the resummation (\ref{dres2c}), while in the pQCD case Eq.~(\ref{dres2b}) a regularisation was performed in \cite{pPLB}, namely $a(\mu^2) \mapsto {\rm Re} \; a(\mu^2 + i \varepsilon)$, to avoid the Landau singularities of the pQCD coupling.\footnote{We called this regularisation the Principal Value (PV) in \cite{pPLB} because it has some (limited) similarity to the PV-regularisation in the Borel resummation.}

In the resummations (\ref{dres2b})-(\ref{dres2c}), $G_d(t)$ is the characteristic function of the canonical BSR $d(Q^2)$ \cite{renmod,pPLB,ACT2023,Castro}
\be
G_d(t) =   \Theta(1-t) \pi \left[ \frac{{\td}^{\rm IR}_1 t}{\Gamma(1-{k_1}) \ln^{{k_1}}(1/t)} + {\td}^{\rm IR}_2 t^2 \right]
+ \Theta(t-1) \pi \left[ \frac{{\td}^{\rm UV}_1}{t} + \frac{{\td}^{\rm UV}_2}{t^2} \right].
\label{Gd} \ee
Here, the values of the residue parameters ${\td}^{\rm IR}_j$ and ${\td}^{\rm UV}_j$ ($j=1,2$), and of the rescaling parameter $\tK$ in Eqs.~(\ref{dres2b})-(\ref{dres2c}), are determined by the knowledge of the first five coefficients of the power expansion (\ref{dptkap1}): $d_0(=1)$, $d_1, \ldots, d_4$. Since the latter coefficients depend on the choice of the renormalisation scheme, so do the residue parameters and $\tK$. We refer for all the details to \cite{pPLB}. We note that the first two terms in $G_d(t)$ come from the first two IR renormalons ($u=1,2$), and the other two terms from the UV renormalons ($u=-1,-2$).\footnote{In Refs.~\cite{Pin1,Pin2} it was argued that the dominance of the $u=1$ renormalon (IR1) (in the $\MSbar$ scheme) in BSR gives a good prediction of the known coefficent $d_3$.}

The (massless) renormalisation scheme is determined by the coefficients $c_2$ and $c_3$ that appear in the renormalisation group equation (RGE)
\be
\frac{d {a}(\mu^2)}{d \ln \mu^2} =- \beta_0 a(\mu^2)^2 \left[ 1 + c_1 a(\mu^2) + c_2 a(\mu^2)^2 + c_3 a(\mu^2)^3 + \ldots \right].
\label{RGE2} \ee
In fact, the scheme is determined by the entire set of the coefficients $c_j$ ($j \geq 2$) \cite{Stevenson}, but we use a set of beta-functions of a specific form\footnote{A Pad\'e form $\beta(a)=[4/4](a)$, so we call this class of schemes P44.}
which has only $c_2$ and $c_3$ parameters freely adjustable \cite{pPLB} and conveniently allows for an explicit solution of the pQCD running coupling $a(\mu^2)$ in terms of the Lambert function \cite{GCIK}. We vary the scheme parameters, as explained and motivated in \cite{pPLB}, in a specific range
\be
c_2 = 9^{+2}_{-1.4}, \quad c_3= 20 \pm 15,
\label{c2c3} \ee
primarily in order to avoid numerical instabilities coming from strong cancellations of the the infrared (IR) renormalon contributions to the resummed value of $d(Q^2)$. The scheme for holomorphic (AQCD) couplings $\A(\mu^2)$ refers to the scheme of the underlying pQCD coupling $a(\mu^2)$, cf.~Appendix \ref{app:AQCD}.

\section{Evaluation of the canonical part $d(Q^2)$}
\label{sec:AQCDcan}

\begin{figure}[htb] 
\centering\includegraphics[width=80mm]{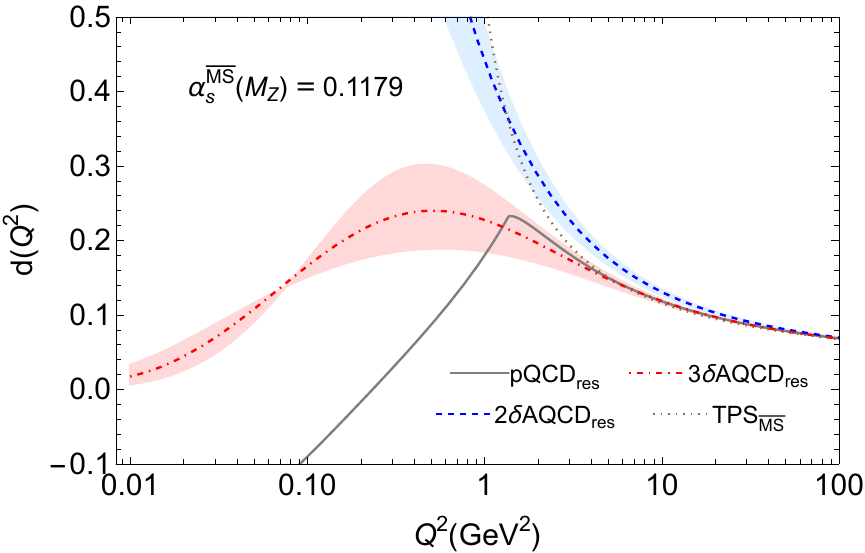}
 \vspace{4pt}
 \caption{\footnotesize The resummed canonical part of BSR, $d(Q^2)_{\rm res}$ (at $N_f=3$), according to Eq.~(\ref{dres2c}), for the 2$\delta$ACD and 3$\delta$AQCD coupling ('$2\delta{\rm AQCD}_{\rm res}$', '$3\delta{\rm AQCD}_{\rm res}$'), with three hadronic threshold scales $M_1$ of their spectral function. For comparison, the resummed result Eq.~(\ref{dres2b}) is also shown, for the pQCD coupling that is the underlying pQCD coupling of 2$\delta$ACD ('${\rm pQCD}_{\rm res}$'). Also included is the truncated perturbation series (TPS) in powers of $a=a(Q^2)$ in the $\MSbar$ scheme up to $\sim a^5$ ('${\rm TPS}_{\MSbar}$'). All involved couplings correspond to the strength $\alpha_s^{\MSbar}(M_Z^2;N_f=5)=0.1179$, and the $d_4$ coefficient value corresponds to the central value of the corresponding $\MSbar$ coefficient, $d_4^{\MSbar}=1557.4$. See the text for more details.}
\label{FigdQ2}
\end{figure}

In Fig.~\ref{FigdQ2} we present these resummed values $d(Q^2)_{\rm res}$ as a function of $Q^2$ [Eq.~(\ref{dres2c})], when either $3\delta$AQCD or $2\delta$AQCD holomorphic coupling is used. The value of the Lambert scale of the coupling is taken to be $\Lambda_{L}=0.21745$ GeV, this corresponds to the value\footnote{For details of the $N_f=3$ explicit expression $a(\mu^2)$ in the P44 renormalisation scheme with $c_2$ and $c_3$ values, and how this is related to the $N_f=5$ value $\alpha_s(M_Z^2)$ in the $\MSbar$ schemes, we refer to \cite{pPLB} and additional references cited there.}
$\alpha_s^{\MSbar}(M_Z^2)=0.1179$.
The QCD variant $2 \delta$AQCD is in the P44-scheme with $c_2=9.$ and $c_3=20.$, and the variant $3 \delta$AQCD is in the (lattice-related) Lambert MiniMOM (LMM) P44-scheme \cite{MM1,MM2a,MM2b,MM3}.\footnote{The (P44) LMM scheme ($c_2^{\rm LMM} =9.29703$ and $c_3 ^{\rm LMM} =71.4538$) is tied to 3$\delta$AQCD and must be used there, cf.~Appendix \ref{app:AQCD}.} The construction of these QCD variants is explained in Appendix \ref{app:AQCD}, and they are used also in the numerical fits later on in Sec.~\ref{sec:fit}. In both these holomorphic cases the strength of the underlying pQCD coupling $a(Q^{'2})$ corresponds to $\alpha_s^{\MSbar}(M_Z^2)=0.1179$, and the threshold mass $M_1$ of the spectral function of the coupling $\A(Q^{'2})$ [cf.~Eq.~(\ref{rhoAnd}) and its description there] has the chosen values $M_1=0.100,0.150$,$0.250$ GeV (the bands in the Figure correspond to the variation $M_1=0.150^{+0.100}_{-0.050}$ GeV). We refer to Appendix \ref{app:AQCD} for more details. For comparison, we also include the resummed values when pQCD coupling is used, Eq.~(\ref{dres2b}), and that coupling is in the  P44-scheme with $c_2=9.$ and $c_3=20$. Furthermore, the values of the simple truncated perturbation series (TPS, truncated at $\sim a^5$) in $\MSbar$ scheme are given.\footnote{The resummed pQCD curve, the TPS curve, and the resummed $3\delta$AQCD curve for $M_1=0.150$ GeV were already given in Fig.~1 of Ref.~\cite{pPLB}.}

As seen in Fig.~\ref{FigdQ2}, the curve for the pQCD case has a (soft) kink at $Q^2 \approx   1.44 \ {\rm GeV}^2$. Such kinks do not appear in the cases of AQCD. The kink comes from the Landau singularities of the pQCD coupling $a(t e^{- \tK} Q^2+ i \epsilon)$ in the integrand of the resummation (\ref{dres2b}) at low $t Q^2$, and the effect of these singularities becomes rather abruptly more pronounced when $Q^2$ has lower values ($Q^2 \leq 1.44 \ {\rm GeV}^2$). Furthermore, we see in Fig.~\ref{FigdQ2} that the curve of $d(Q^2)$ for 2$\delta$AQCD converges to the asymptotic (pQCD) behaviour at quite high $Q^2$ (in contrast to the 3$\delta$AQCD curve); this is related to the fact that the 2$\delta$AQCD coupling $\A(Q^{'2})$ at low $Q^{'2}$ reaches relatively high values (cf.~Appendix \ref{app:AQCD}) and these contributions are significant in the resummation integral (\ref{dres2c}) at low $t$ values, even when $Q^2$ is relatively high. We remark that in Fig.~\ref{FigdQ2} the pQCD TPS curve (in $\MSbar$, with $\kappa=1$) becomes infinite at $Q^2 \approx 0.40 \ {\rm GeV}^2$, which is the branching point of the Landau cut of the $\MSbar$ pQCD coupling; for $Q^2 < 0.40 \ {\rm GeV}^2$, that TPS curve is negative and thus unphysical. On the other hand, the 2$\delta$AQCD and 3$\delta$AQCD curves in Fig.~\ref{FigdQ2} remain finite and positive all the way down to $Q^2=0$ where they reach the values of $3.70$ (when $M_1=0.150$ GeV) and zero, respectively.
\begin{figure}[htb] 
\begin{minipage}[b]{.49\linewidth}
\includegraphics[width=80mm,height=50mm]{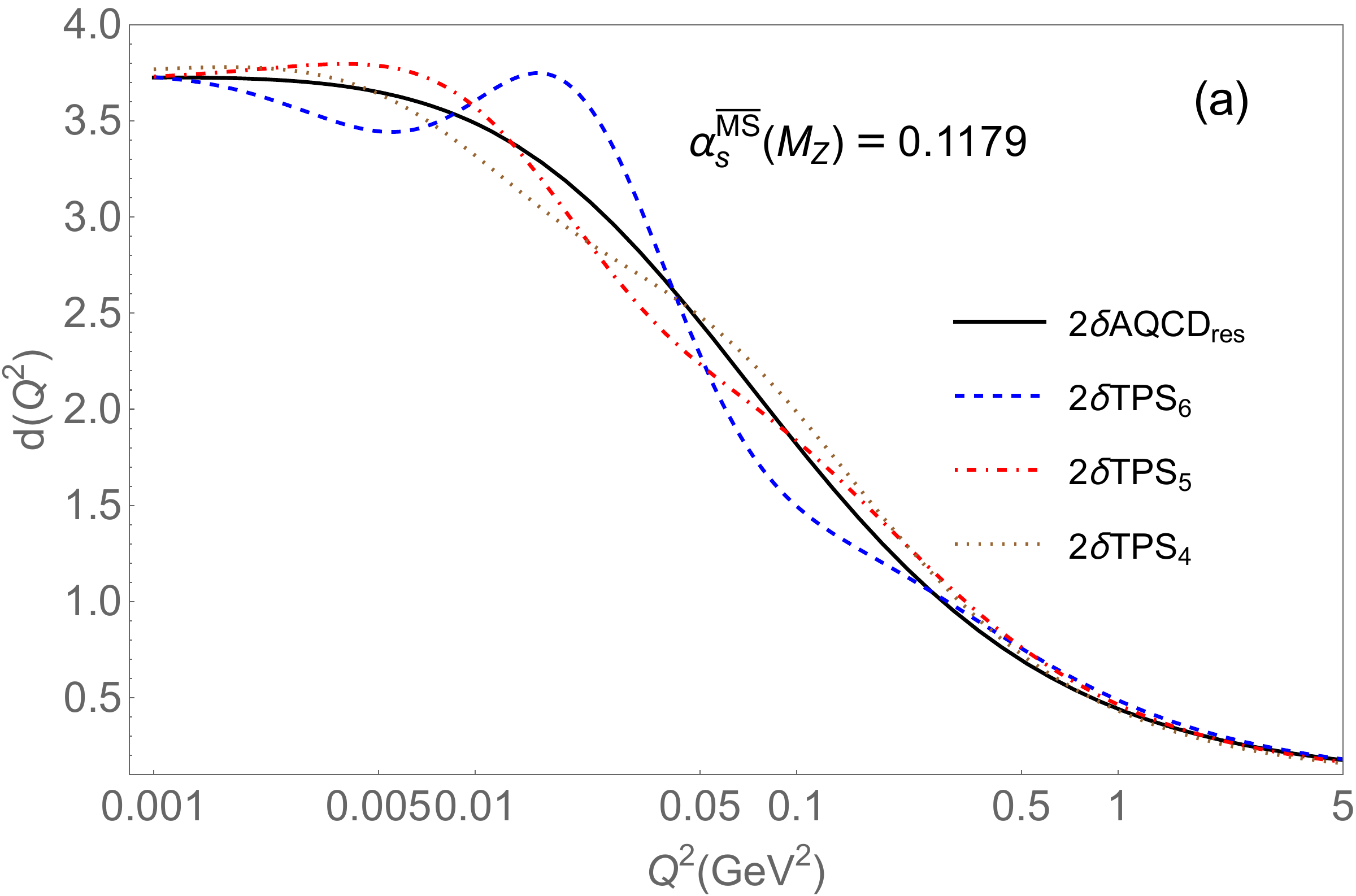}
\end{minipage}
\begin{minipage}[b]{.49\linewidth}
  \includegraphics[width=80mm,height=50mm]{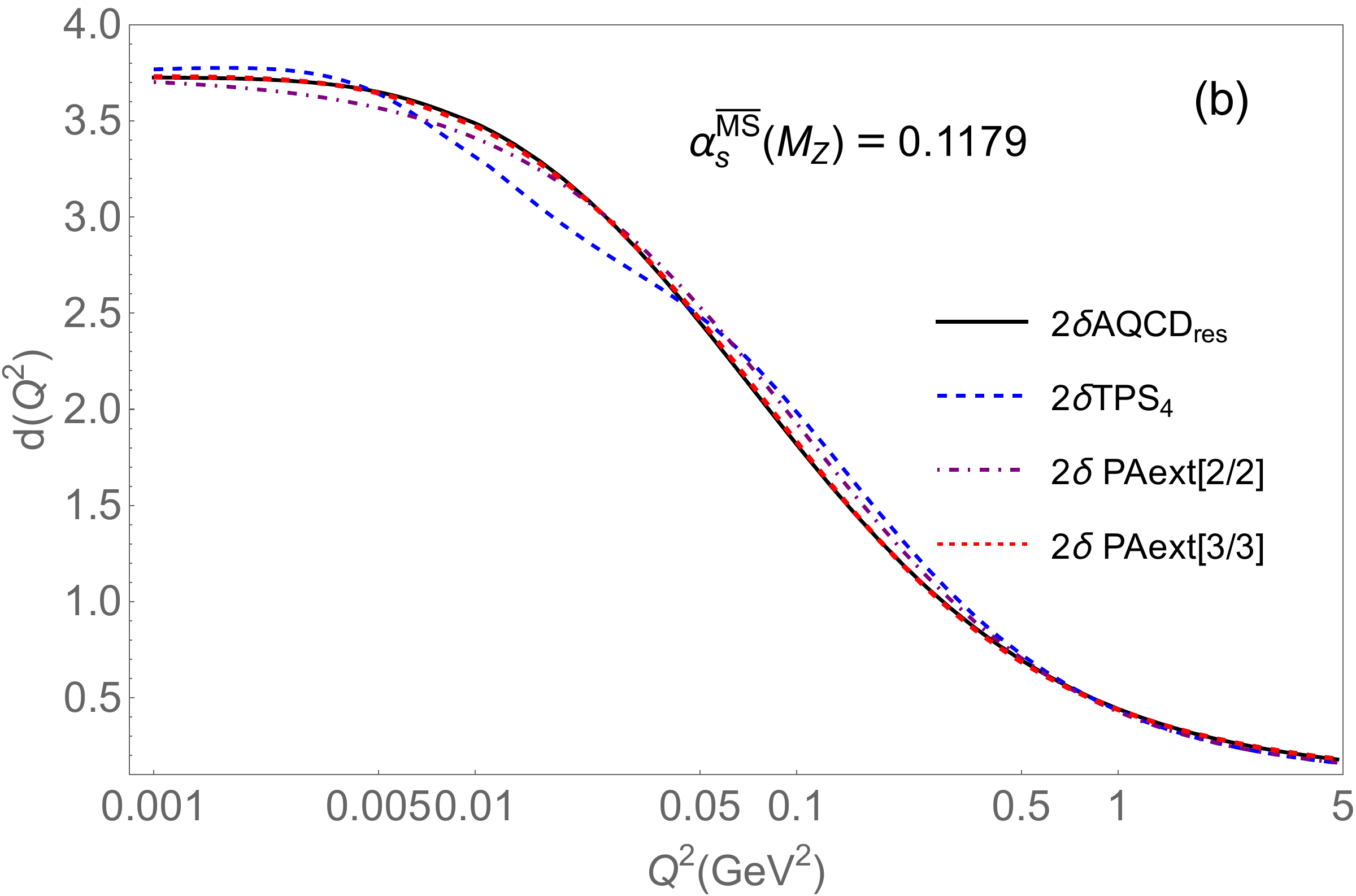}
\end{minipage} \vspace{4pt}
\caption{\footnotesize The resummed $d(Q^2)$ in 2$\delta$AQCD, compared with various approximants based on truncated information about the coefficients $\td_n$ ($n \leq 5$): (a) the truncated series $d(Q^2)^{[N]}_{2\delta{\rm AQCD}}$ (\ref{dlAptTr}) ('$2\delta{\rm TPS}_{N}$') for $N=4,5,6$; (b) the Pad\'e-related renormalisation scale-invariant approximants  ${\cal G}_d^{[M/M]}(Q^2)_{2\delta{\rm AQCD}}$ Eq.~(\ref{GMM}) ('$2\delta$PAext[M/M]'), for $M=2$ and $M=3$. Input parameters for the coupling are $\alpha_s^{\rm MS}(M_Z^2)=0.1179$ and $M_1=0.150$ GeV. The value of the coefficient $d_4$ corresponds to $d_4^{\MSbar}=1557.4$. We note that the $2\delta$PAext[3/3] curve practically (visually) coincides with the resummed curve.}
\label{FigdQ22d}
\end{figure}
\begin{figure}[htb] 
\begin{minipage}[b]{.49\linewidth}
\includegraphics[width=80mm,height=50mm]{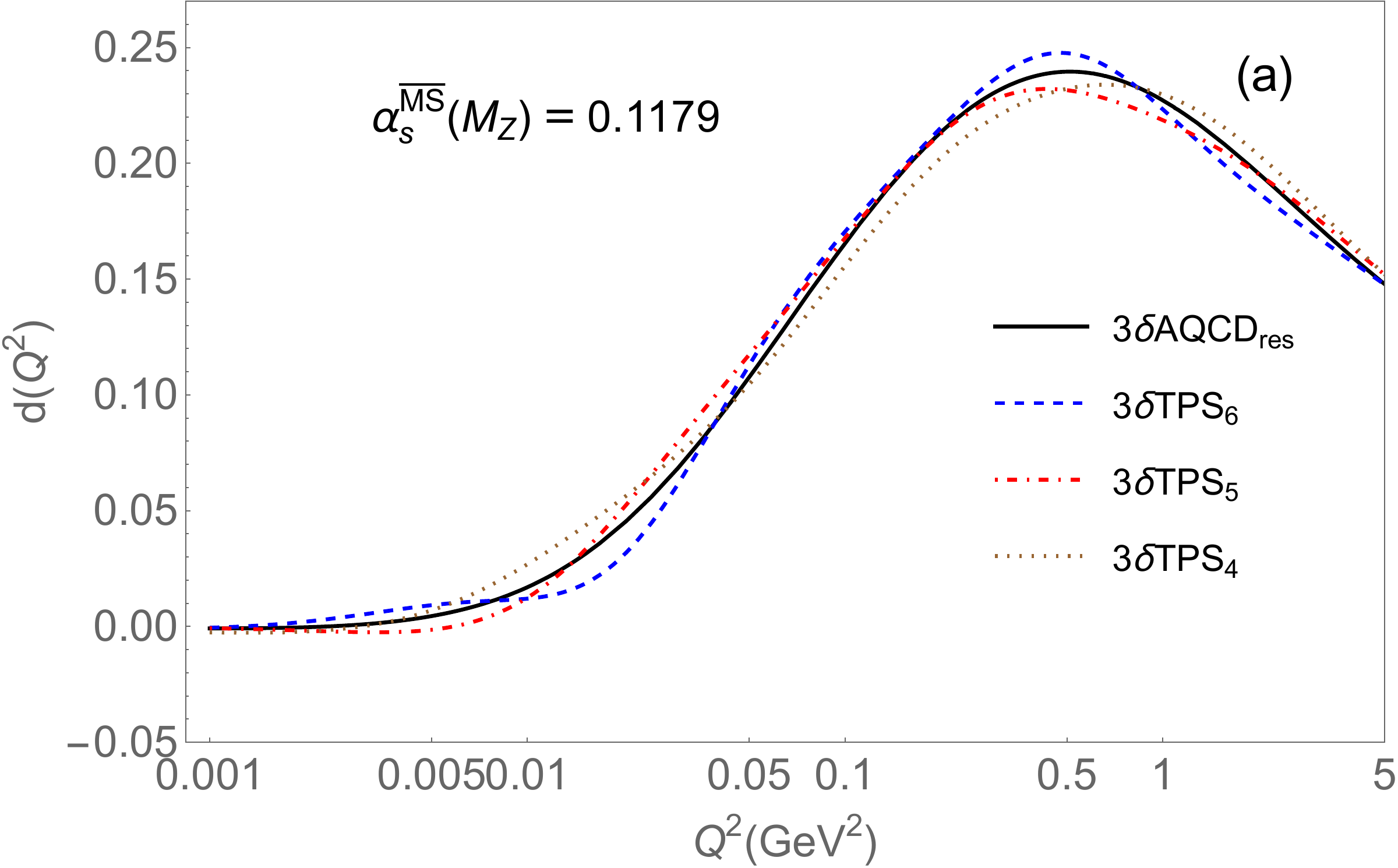}
\end{minipage}
\begin{minipage}[b]{.49\linewidth}
  \includegraphics[width=80mm,height=50mm]{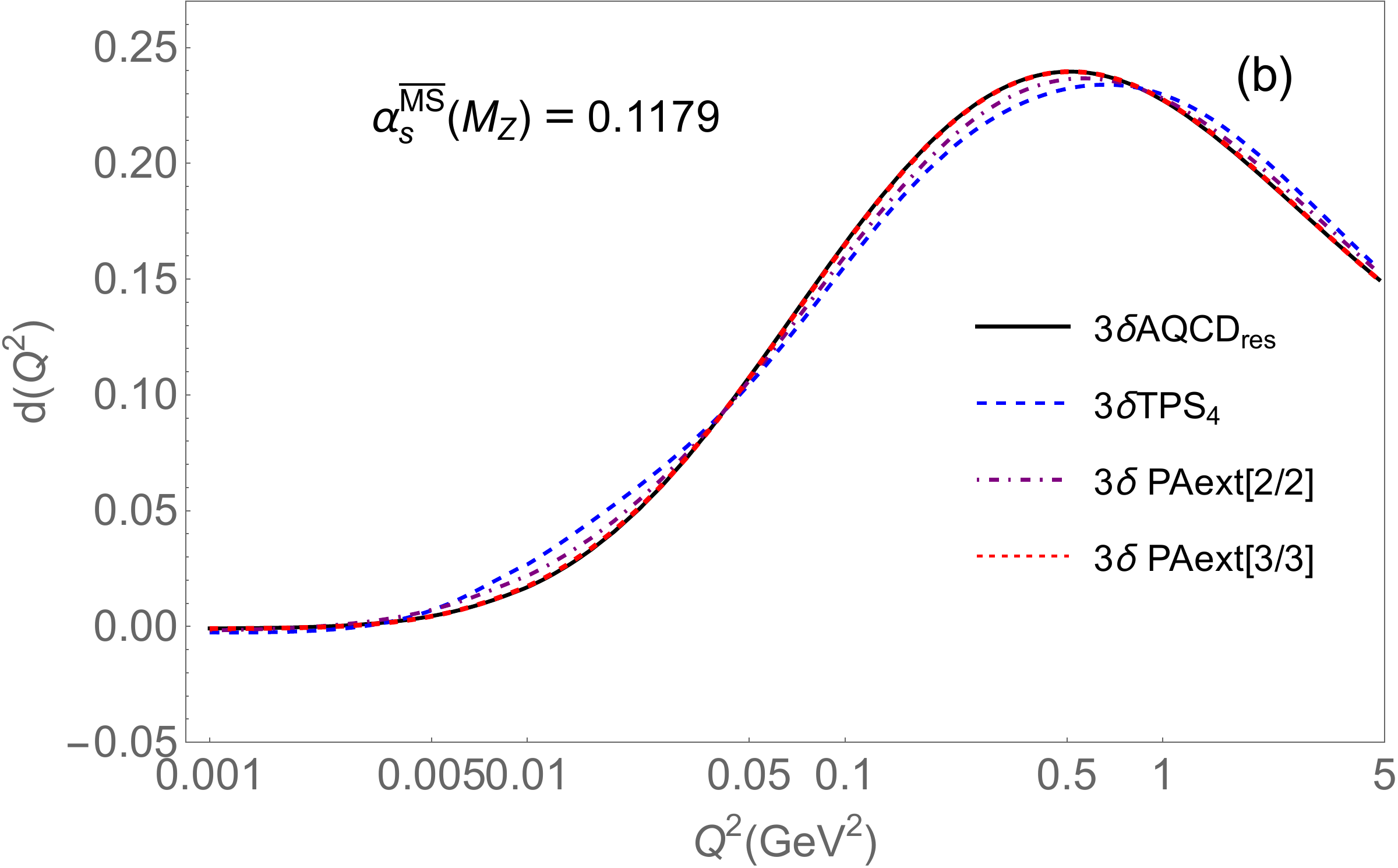}
\end{minipage} \vspace{4pt}
\caption{\footnotesize The same as in Figs.~\ref{FigdQ22d}, but in 3$\delta$AQCD (instead of 2$\delta$AQCD). We note that the $3\delta$PAext[3/3] curve practically (visually) coincides with the resummed curve. }
\label{FigdQ23d}
\end{figure}

We present in Figs.~\ref{FigdQ22d}(a) and \ref{FigdQ23d}(a) the results for $d(Q^2)$ in  2$\delta$AQCD and 3$\delta$AQCD in the resummed form (\ref{dres2c}) and compare them to the truncated ``perturbation'' series in logarithmic derivatives in those models, i.e., AQCD analogues of the pQCD series Eq.~(\ref{dlpt}), as presented in Appendix \ref{app:AQCD}, Eqs.~(\ref{tAn})-(\ref{dlApt}), with $\kappa=1$ and truncated at $\sim \tA_N$
\be
d(Q^2)_{\rm AQCD}^{[N]} = \A(Q^2) + {\td}_1 \; {{\tA}}_2(Q^2) + \ldots + {\td}_{N-1} \; {{\tA}}_{N}(Q^2),
\label{dlAptTr} \ee
where we denoted as earlier, for simplicity, $\td_n(\kappa=1) \equiv \td_n$.
We note that formally, $\tA_N \sim a^N$ where $a$ is the underlying pQCD coupling. We can observe in Fig.~\ref{FigdQ22d}(a), and to a lesser degree in \ref{FigdQ23d}(a), numerical indications that the truncated AQCD series (\ref{dlAptTr}) give us a divergent sequence when the truncation index $N$ increases (in the figures we have $4 \leq N \leq 6$), i.e., we have no convergence to the resummed curves when $N$ increases.

On the other hand, in Figs.~\ref{FigdQ22d}(b) and \ref{FigdQ23d}(b) we include the curves of the Pad\'e-related approximants  ${\cal G}_d^{[M/M]}(Q^2)_{{\rm AQCD}}$ (denoted as 'PAext[M/M]': 'extensions' of (diagonal) Pad\'e's). The approximants were introduced in the framework of pQCD in \cite{dBGpt}. They were later applied in QCD variants with holomorphic coupling (AQCD, i.e., free of Landau singularities) in Refs.~\cite{dBG1,dBG2,dBG3,renmod}; they were applied there to the (spacelike) Adler function and to related QCD observables. These approximants are briefly explained in Appendix \ref{app:AQCD} [Eqs.~(\ref{GMM})-(\ref{GMMprec})].
These approximants are constructed only from the first $N=2 M$ expansion coefficients $\td_n(\kappa)$ of $d(Q^2)$ [i.e., from the coefficients $\td_j(\kappa)$, $0 \leq j \leq (2 M-1)$], they are entirely independent of the renormalisation scale parameter $\kappa$, and they formally approximate $d(Q^2)$ up to the precision $\sim \tA_{2 M+1}$, i.e., the difference $[d(Q^2) - {\cal G}_d^{[M/M]}(Q^2)_{{\rm AQCD}}]$ is $\sim \tA_{2 M+1}$ (and this is formally $\sim a^{2 M+1}$, where $a$ is the underlying pQCD coupling).\footnote{
We point out that these approximants ${\cal G}_d^{[M/M]}(Q^2)$ often cannot be applied in pQCD in practice because of the presence of Landau poles in $a(Q^{' 2})$ at low positive $Q^{2 '}$. In this respect, we note these approximants contain pQCD terms $\tal_j a(\kappa_j Q^2)$ for various $\tal_j$ and $\kappa_j>0$, and usually some $\kappa_j$'s have low values ($\kappa_j \ll 1$), cf.~Table \ref{tabdBG} in Appendix \ref{app:AQCD}.}
  We can see in Figs.~\ref{FigdQ22d}(b) and \ref{FigdQ23d}(b) that these approximants converge quickly toward the resummed value of $d(Q^2)$, Eq.~(\ref{dres2c}), when $N$($=2 M$) increases. This is in stark contrast to the corresponding truncated series approximants (\ref{dlAptTr}) with $N = 2 M$, cf.~Figs.~\ref{FigdQ22d}(a) and \ref{FigdQ23d}(a).

\section{Fits to the experimental data, in AQCD}
\label{sec:fit}

The experimental data for inelastic BSR ${\overline \Gamma}_1^{\rm p-n}(Q^2)$ have been obtained by various experiments \cite{CERN,DESY,SLAC,JeffL1,JeffL2,JeffL3,JeffL4,JeffL5},  for the range $0<Q^2< 10 \ {\rm GeV}^2$. Our compilation of these data, for $0 < Q^2 \leq 4.739 \ {\rm GeV}^2$, with statistical and systematic uncertainties, is presented in Figs.~2 of our previous work \cite{pPLB}, and will be used here as well. We point out that we use in our approach $N_f=3$ throughout. There are some experimental data available also for $Q^2 > 4.739 \ {\rm GeV}^2$, but we will not take them into account, because that would require adjusting our resummation formalism to the $N_f=4$ regime. This adjustment, in our resummation approach and with our AQCD couplings, is nontrivial and we have not performed it. Furthermore, the data at such high $Q^2$ are not many and have, in general, larger statistical and systematic uncertainties.

In this Section we will present the fit results with the approach Eq.~(\ref{dres2c}) using the 2$\delta$AQCD and 3$\delta$AQCD couplings, and compare them with the same approach using pQCD coupling (in the P44 scheme with  $c_2=9.$ and $c_3=20.$) that were obtained in our previous work \cite{pPLB}.

We perform the fit where the above experimental values are fitted with the theoretical OPE expresssion Eq.~(\ref{BSROPE}) truncated either at $D=2$, or at $D=4$, and where the QCD canonical part $d(Q^2)$ is evaluated with the renormalon-based resummation Eq.~(\ref{dres2c}).

\begin{figure}[htb] 
\begin{minipage}[b]{.49\linewidth}
\includegraphics[width=80mm,height=50mm]{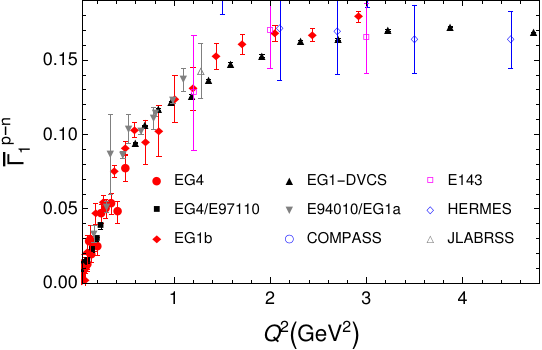}
\end{minipage}
\begin{minipage}[b]{.49\linewidth}
  \includegraphics[width=80mm,height=50mm]{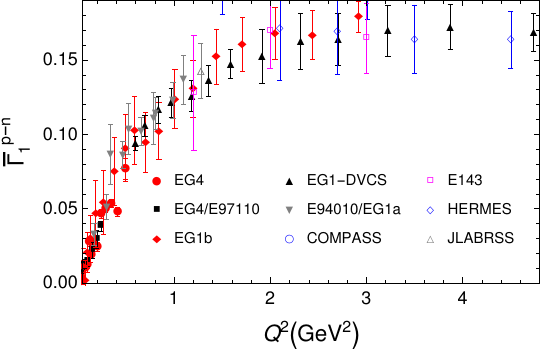}
\end{minipage} \vspace{12pt}
\caption{\footnotesize The measured results for the inelastic BSR ${\overline \Gamma}_1^{p-n}(Q^2)$ for different experiments, with the statistical (left Figure) and systematic (right Figure) uncertainties. The Figures were taken from \cite{pPLB}.}
\label{Figstasys}
\end{figure}

The fit consists of varying either the one fit parameter ${\bar f}_2$, or the two fit parameters ${\bar f}_2$ and $\mu_6$, cf.~Eq.~(\ref{BSROPE}), such that the quantity 
\be
\chi^2(j_{\rm min}; k) = \frac{1}{(j_{\rm max}-j_{\rm min}+1-N_{\rm p})}\sum_{j=j_{\rm min}}^{j_{\rm max}} \frac{ \left[ {\overline \Gamma}_1^{{\rm p-n},{\rm OPE}}(Q_j^2) - {\overline \Gamma}_1^{{\rm p-n}}(Q_j^2)_{\rm exp} \right]^2}{\sigma(Q_j^2; k)^2}
\label{chi2} \ee
is minimised. Here, $Q^2_j$ (where: $j=1,\ldots,j_{\rm max}$; $j_{\rm max}=77$) are the squared momentum scales at which the experimental data are available $(Q_1^2 < Q_2^2 < \cdots < Q_{77}^2)$. Further, $N_{\rm p}$ is the number of fit parameters: $N_{\rm p}=1$ if $\mu_6=0$ and ${\bar f}_2$ is varied; $N_{\rm p}=2$ if both ${\bar f}_2$ and $\mu_6$ are varied. The index parameter $j_{\rm min}$ indicates a chosen minimal $Q^2_{\rm min}=Q^2_{j_{\rm min}}$ scale for which the experimental data are included in the fit; and $j_{\rm max}=77$ corresponds to the maximal fit scale, $Q^2_{\rm max}=Q_{77}^2=4.739 \ {\rm GeV}^2$. Hence, the interval of scales included in the fit is: $Q^2_{\rm j_{\rm min}} \leq Q^2 \leq 4.739 \ {\rm GeV}^2$. In Figs.~\ref{Figstasys} we present, for convenience, our compilation \cite{pPLB} of the measured BSR values ${\overline \Gamma}_1^{{\rm p-n}}(Q_j^2)_{\rm exp}$ from various experiments, with statistical and systematic uncertainties, for $Q_j^2 \leq 4.739 \ {\rm GeV}^2$.
 The uncorrelated squared uncertainties $\sigma(Q_j^2; k)^2$ at $Q_j^2$ in the expression (\ref{chi2}) are in principle unknown. The statistical errors $\sigma_{\rm stat}(Q_j^2)$ are uncorrelated, but the systematic errors $\sigma_{\rm sys}(Q_j^2)$ may have significant, but unknown, correlations. Therefore, we follow here the method of unbiased estimate \cite{Deuretal2022,PDG2020,Schmell1995}. This method consists of the following.
A fraction of $\sigma^2_{\rm sys}(Q_j^2)$ is added to $\sigma^2_{\rm stat}(Q_j^2)$
\be
\sigma^2(Q_j^2; k) = \sigma^2_{\rm stat}(Q_j^2) + k \; \sigma^2_{\rm sys}(Q_j^2).
\label{sig2} \ee
The obtained uncertainties $\sigma(Q_j^2; k)$ are regarded as uncorrelated, and the mentioned fit parameters (only ${\bar f}_2$; or ${\bar f}_2$ and $\mu_6$) are extracted by minimisation of the above expression $\chi^2(j_{\rm min};k)$, Eq.~(\ref{chi2}), in the mentioned chosen interval of $Q^2$ values. This process is continued, by adjusting iteratively the parameter $k$ and minimising $\chi^2(j_{\rm min};k)$ until the value $\chi^2(j_{\rm min};k)=1$ is obtained. In practice, in this way we always obtain $0 < k < 0.5$. We note that the smaller the obtained value of $k$, the more precise is the fit.

The experimental uncorrelated uncertainty (exp.u.) of the obtained fit parameters (${\bar f}_2$; or ${\bar f}_2$ and $\mu_6$) is then obtained by the conventional method as explained, e.g., in App. of Ref.~\cite{Bo2011}, or App.~D of \cite{ACT2023}. For completeness, we describe this method of obtaining 'exp.u.' in Appendix \ref{app:expu}.

The experimental correlated uncertainty (exp.c.) is then obtained by simply shifting the central experimental values ${\overline \Gamma}_1^{{\rm p-n}}(Q_j^2)_{\rm exp}$ in the expression (\ref{chi2}) by the errors complementary to $\sigma(Q_j^2; k)$ of Eq.~(\ref{sig2}), namely by $(1 - \sqrt{k}) \sigma_{\rm sys}(Q_j^2)$, up and down, and conducting the minimisation of this new $\chi^2(j_{\rm min}; k)$. The corresponding variation '(exp.c.)' of the extracted parameters is then the difference between such ``shifted'' extracted values and the central (``unshifted'') values.

Another question is how to choose the preferred value of $Q^2_{\rm min}$ ($=Q^2_{j_{\rm min}}$). The results of the fit can depend considerably on the choice of the value of $Q^2_{\rm min}$.
In pQCD, the choice was $Q^2_{\rm min}=1.71 \ {\rm GeV}^2$ in the fit with either two parameters (${\bar f}_2$, $\mu_6$) or one parameter (${\bar f}_2$), we refer for details to \cite{pPLB}.\footnote{In the used scheme P44 with $c_2=9.$ and $c_3=20.$ (and $N_f=3$), and with the strength scale $\Lambda_L=0.2175$ GeV corresponding to $\alpha_s^{\MSbar}(M_Z^2)=0.1179$, the Landau cut of $a(Q^{'2})$ is $0 \leq Q^{'2} < 0.87 \ {\rm GeV}^2$.}
We discuss the determination of the values of $Q^2_{\rm min}$ in our AQCD analyses later on in this Section.

Yet another uncertainty of the extracted values of parameters, related to the experimental statistical and systematic uncertainties of the measured BSR values, is the ('$k$') uncertainty originating from the different ways of fixing the $k$ value of Eq.~(\ref{sig2}) for the fits. As mentioned, we determined one common value of $k$ for all the experimental BSR data (but for each fit analysis separately). However, the experimental data used (in the $N_f=3$ regime $0.5 \ {\rm GeV}^2 < Q^2 \leq 4.739 \ {\rm GeV}^2$) belong to different experiments, so that for each experiment we have a different value of $k$. We addressed this problem in Appendix \ref{app:ks}, where we show that by far the largest number of points is from one experiment (Hall B of Jefferson Lab), and it is the value of $k$ from that experiment that dominates the fit analysis (especially because the results of other experiments have in general larger uncertainties). We also show that the extracted values of parameters change only moderately if we apply this multi-$k$ approach to the two-parameter fits, and that the changes are insignificant in the one-parameter fits. These changes are indicated as ('$k$') uncertainties in the values of the extracted parameters below.

As mentioned, in the present resummation approach with AQCD couplings, Eq.~(\ref{dres2c}),
 the employed variants are 2$\delta$AQCD \cite{2dAQCD,2dAQCDb} and 3$\delta$AQCD \cite{3dAQCD,3dAQCDb}, which are briefly described here in Appendix \ref{app:AQCD} and were mentioned in Sec.~\ref{sec:AQCDcan}. The renormalisation schemes for 2$\delta$AQCD coupling are again taken to be the P44-schemes with the $c_2$ and $c_3$ beta-parameters varying in the range (\ref{c2c3}) as in pQCD; i.e., the central case will be again $c_2=9.$ and $c_3=20$. On the other hand, the 3$\delta$QCD coupling is related to the large volume lattice results \cite{BIMS1,BIMS2} and is thus in a fixed lattice-related scheme, the Lambert MiniMOM (LMM) P44-scheme \cite{MM1,MM2a,MM2b,MM3} ($c_2^{\rm LMM} =9.2980$ and $c_3 ^{\rm LMM} =71.4538$). As explained in Appendix \ref{app:AQCD}, in each of the two mentioned AQCD couplings, the parameters that fix the coupling are two: (a) the value of $\alpha_s^{\MSbar}(M_Z^2)$ that determines the ($N_f=3$) scale $\Lambda_L$ and thus determines the underlying $N_f=3$ pQCD coupling; (b) the scale (in GeV) of the lowest threshold mass $M_1=\sqrt{\sigma_{\rm min}}$ of the spectral (discontinuity) function $\rho(\sigma) = {\rm Im} \A(-\sigma - i \epsilon)$. The threshold scale $M_1$ is expected to be of the order of the lowest hadronic scale, $M_1 \sim m_{\pi} \approx 0.150$ GeV. So, for the central case, we fix the threshold scale to the value $M_1 = 0.150$ GeV and take $\alpha_s^{\MSbar}(M_Z^2)=0.1179$.

The fitting procedure gives for the 2$\delta$AQCD coupling for the two-parameter fit the values $k=0.1342$ and
\bes
\label{resLTD242d}
\bea
{\bar f}_2^{(2 \delta)} & = & +0.0041^{+0.0151}_{-0.0109}(c_2)^{+0.0083}_{-0.0148}(c_3)^{+0.0489}_{-0.0466}(\alpha_s) \pm 0.0034(d_4)^{-0.0770}_{+0.0557}(M_1)^{-0.0129}_{+0.0128}(a_2d_2)
\nonumber\\ &&
^{+0.0388}_{-0.0238}(Q^2_{\rm min}) \pm 0.0233({\rm exp.u.}) \pm 0.1314({\rm exp.c.}) \mp 0.0125(k),
\label{bf2D242d} \\
\mu_6^{(2 \delta)} & = & -0.0016 \mp 0.0004(c_2)^{+0.0006}_{-0.0004}(c_3)^{-0.0029}_{+0.0025}(\alpha_s) \mp 0.0005(d_4)^{+0.0001}_{-0.0002}(M_1) \mp 0.0011(a_2d_2)
\nonumber\\ &&
^{-0.0076}_{+0.0035}(Q^2_{\rm min}) \pm 0.0031({\rm exp.u.}) \mp 0.0148({\rm exp.c.}) \pm 0.0016(k)
\; [{\rm GeV}^4].
\label{mu6D242d} \eea \ees
For the one-parameter fit with $2\delta$AQCD we obtain the values $k=0.1295$ and 
\bea
{\bar f}_2^{(2 \delta)} & = & -0.0082^{+0.0124}_{-0.0087}(c_2)^{+0.0130}_{-0.0176}(c_3)^{+0.0283}_{-0.0274}(\alpha_s)^{-0.0001}_{+0.0003}(d_4)^{-0.0763}_{+0.0541}(M_1)^{-0.0209}_{+0.0210}(a_2d_2)
\nonumber\\ &&
^{-0.0006}_{+0.0035}(Q^2_{\rm min}) \pm 0.0038({\rm exp.u.}) \pm 0.0207({\rm exp.c.}) \pm 0.0001(k).
\label{bf2D22d} \eea

For the 3$\delta$AQCD coupling for the two-parameter fit we obtain the values $k=0.1657$ and
\bes
\label{resLTD243d}
\bea
{\bar f}_2^{(3 \delta)} & = & -0.2488^{+0.0244}_{-0.0237}(\alpha_s)^{-0.0014}_{+0.0013}(d_4)^{+0.0781}_{-0.0823}(M_1)^{-0.0148}_{+0.0149}(a_2d_2)
\nonumber\\ &&
^{+0.0229}_{-0.0132}(Q^2_{\rm min}) \pm 0.0247({\rm exp.u.}) \pm 0.1193({\rm exp.c.}) \mp 0.0201(k),
\label{bf2D243d} \\
\mu_6^{(3 \delta)} & = & -0.0006^{-0.0018}_{+0.0017}(\alpha_s) \pm 0.0002(d_4)^{-0.0045}_{+0.0051}(M_1) \mp 0.0005 (a_2d_2)
\nonumber\\ &&
^{-0.0040}_{+0.0020}(Q^2_{\rm min}) \pm 0.0032({\rm exp.u.}) \mp 0.0131({\rm exp.c.}) \pm 0.0025(k)
\; [{\rm GeV}^4].
\label{mu6D243d} \eea \ees
For the one-parameter fit with $3\delta$AQCD we obtain the values $k=0.1583$ and 
\bea
{\bar f}_2^{(3 \delta)} & = & -0.2534^{+0.0110}_{-0.0109}(\alpha_s) \pm 0.0001(d_4)^{+0.0447}_{-0.0427}(M_1)^{-0.0187}_{+0.0186}(a_2d_2)
\nonumber\\ &&
^{-0.0044}_{+0.0019}(Q^2_{\rm min}) \pm 0.0042({\rm exp.u.}) \pm 0.0202({\rm exp.c.}) \mp 0.0013(k).
\label{bf2D23d} \eea
The parameter ${\bar f}_2$ that appears in the $D=2$ OPE term Eq.~(\ref{BSROPE}) is dimensionless, but the $D=4$ OPE parameter $\mu_6$ is in units of ${\rm GeV}^4$.

As mentioned above, for the 3$\delta$AQCD case we cannot present scheme uncertainties ('$c_2$') and ('$c_3$') because the construction of the 3$\delta$AQCD coupling is tied to the lattice (L)MM scheme. In the results (\ref{resLTD242d})-(\ref{bf2D22d}), the (theoretical) uncertainties at '$(c_2)$' and '$(c_3)$' originate from the renormalisation scheme variation, Eq.~(\ref{c2c3}). The uncertainty at '$(\alpha_s)$' in all the above results originates from the world average $\alpha_s$-uncertainty $\alpha_s^{\MSbar}(M_Z^2)=0.1179 \pm 0.0009$ \cite{PDG2023}. The (theoretical) uncertainty at ('$d_4$') comes from the mentioned variation  $d_4^{\MSbar}\approx 1557.4 \pm 32.8$.

The (experimental) uncertaintites (exp.u.) and (exp.c) were explained earlier in this Section, as was the ('$k$') uncertainty. We note that the upper sign at each ('$k$') uncertainty brings us from the common-$k$ approach (used here) to the multi-$k$ approach explained in Appendix \ref{app:ks}.
 
In comparison to the pQCD case \cite{pPLB}, we now have no ('ren') uncertainties coming from the renormalon ambiguity, because no regularisation is needed in the resummation (\ref{dres2c}). However, the somewhat analogous (theoretical) uncertainty is the ('$M_1$') uncertainty in Eqs.~(\ref{resLTD242d})-(\ref{bf2D22d}) and (\ref{resLTD243d})-(\ref{bf2D23d}), which comes by varying the mentioned threshold scale $M_1$ of the spectral function $\rho(\sigma)$ of the AQCD coupling, cf.~Eq.~(\ref{rhoAnd}); we performed the following variation of this scale: $M_1=0.150^{+0.100}_{-0.050}$ GeV.

The uncertainty ('$a_2d_2$') comes from the uncertainties of $a_2(Q_0^2)$ and $d_2(Q_0^2)$ at $Q_0^2=4 \ {\rm GeV}^2$ (added in quadrature), cf.~Eq.~(\ref{a2d2A}).

\begin{figure}[htb] 
\begin{minipage}[b]{.49\linewidth}
\includegraphics[width=80mm,height=50mm]{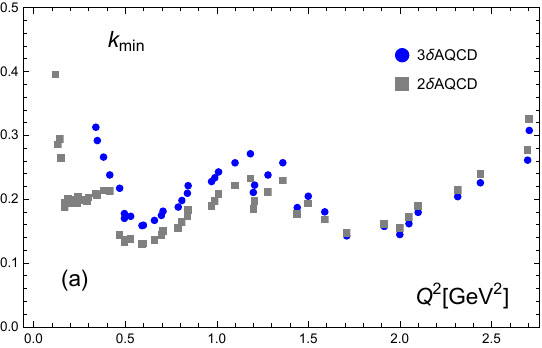}
\end{minipage}
\begin{minipage}[b]{.49\linewidth}
  \includegraphics[width=80mm,height=50mm]{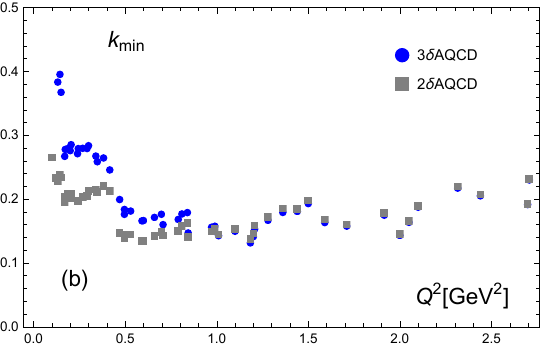}
\end{minipage} \vspace{12pt}
\caption{\footnotesize The values of the $\sigma^2$-parameter $k$, Eq.~(\ref{sig2}), for various $Q^2=Q^2_{\rm min}$ for the 2$\delta$ and 3$\delta$AQCD coupling: (a) when one-parameter fit is performed; (b) when two-parameter fit is performed. }
\label{FigkvsQ2}
\end{figure}

Furthermore, the uncertainty ('$Q^2_{\rm min}$') comes now from the following variation, in the 2$\delta$AQCD case:
\be
(Q^2_{\rm min})^{(2 \delta)} =  0.592^{+0.198}_{-0.122} \; {\rm GeV}^2,
\label{Q2min2d} \ee
which is the same in the two-parameter and one-parameter fit. And in the 3$\delta$AQCD case the variation is
\bes
\label{Q2min3d}
\bea
(Q^2_{\rm min})^{(3 \delta),{\rm 2p}} &=& 0.592^{+0.106}_{-0.096} \; {\rm GeV}^2,
\label{Q2min3d2p} \\
(Q^2_{\rm min})^{(3 \delta),{\rm 1p}} &=& 0.592^{+0.198}_{-0.096} \; {\rm GeV}^2,
\label{Q2min3d1p}
\eea \ees
where in the superscripts '2p' and '1p' mean two-parameter and one-parameter fit, respectively.
The central values for $Q^2_{\rm min}$, in Eqs.~(\ref{Q2min2d})-(\ref{Q2min3d}), were obtained in the following way. For each possible $Q^2_{\rm min}=Q^2_j$
($1 < j < 77$), the fits were performed and the corresponding value of the $\sigma^2$-parameter $k$, Eq.~(\ref{sig2}), was obtained. The results are presented in Figs.~\ref{FigkvsQ2}. The preferred values of $Q^2_{\rm min}$ should not be too close to $Q^2_{\rm max} = Q_{77}=4.739 \ {\rm GeV}^2$, in order to have a reasonably wide $Q^2$-interval for the fitted experimental values. Since the minimal $k$ represents the most precise fit, we choose such $Q^2_{\rm min}$ where the approximately minimal value of $k$ is obtained.\footnote{We could also fix $k$ to a typical value, say $k=0.15$, and then observe $\chi^2(j_{\rm min}; k)$ as a function of $j_{\rm min}$, i.e., as a function of $Q^2_{\rm min}$, and look for such $Q^2_{\rm min}$ where $\chi^2(j_{\rm min}; k)$ is minimal. In this way, in general, we obtain the same or similar value as in the approach described above.}
This gives the central values of $Q^2_{\rm min}$ given above. The variation range of $Q^2_{\rm min}$, as given for each case in Eqs.~(\ref{Q2min2d})-(\ref{Q2min3d}), was then obtained by requiring that beyond the above ranges a relatively abrupt change (increase) in the value of $k$ occurs.

We can compare the obtained results Eqs.~(\ref{resLTD242d}) and  (\ref{resLTD243d}) of the two-parameter fits with the analogous two-parameter fit with pQCD coupling obtained in \cite{pPLB}
\bes
\label{resLTD24}
\bea
{\bar f}_2^{\rm (pQCD)} & = & -0.160^{-0.007}_{+0.025}(c_2)^{+0.054}_{-0.039}(c_3)^{+0.044}_{-0.041}(\alpha_s)^{-0.012}_{+0.016}(d_4) \mp 0.043({\rm ren})
\nonumber\\ &&
^{+0.016}_{+0.119}(Q^2_{\rm min})\pm 0.160({\rm exp.u.}) \pm 0.297({\rm exp.c.}),
\label{bf2D24} \\
\mu_6^{\rm (pQCD)} & = & +0.022^{+0.003}_{-0.008}(c_2)^{-0.013}_{+0.004}(c_3)^{-0.010}_{+0.008}(\alpha_s)^{+0.002}_{-0.003}(d_4) \mp 0.010({\rm ren})
\nonumber\\ &&
^{-0.006}_{-0.053}(Q^2_{\rm min})\pm 0.062({\rm exp.u.}) \mp 0.059({\rm exp.c.})
\; [{\rm GeV}^4],
\label{mu6D24} \eea \ees
and where we had $k=0.1621$ and $Q^2_{\rm min}=1.71^{+0.205}_{-0.27} \ {\rm GeV}^2$.

Further, the results Eq.~(\ref{bf2D22d}) and  (\ref{bf2D23d}) of the one-parameter fits can be compared with the analogous one-parameter fit with pQCD coupling obtained in \cite{pPLB}
\bea
{\bar f}_2^{\rm (pQCD)} & = & -0.107^{-0.001}_{+0.007}(c_2)^{+0.022}_{-0.029}(c_3) \pm 0.020(\alpha_s) \mp 0.009(d_4) \mp 0.067({\rm ren})
\nonumber\\ &&
^{+0.012}_{-0.029}(Q^2_{\rm min}) \pm 0.033({\rm exp.u.}) \pm 0.154({\rm exp.c.}),
\label{bf2D2} \eea
and where we had $k=0.1487$ and $Q^2_{\rm min}=1.71^{+0.39}_{-0.27} \ {\rm GeV}^2$.

We note that these pQCD results were obtained in \cite{pPLB} by using fixed (nonrunning) values of $a_2$ and $d_2$ ($a_2+4 d_2 =0.063$), in contrast to Eq.~(\ref{a2d2p}). However, the central extracted values and the size of uncertainties are only moderately affected by this, especially if $Q^2_{\rm min}$ in the fit is relatively high (as is the case in pQCD: $Q^2_{\rm min}=1.71 \ {\rm GeV}^2$).
\begin{figure}[htb] 
\begin{minipage}[b]{.49\linewidth}
\includegraphics[width=80mm,height=50mm]{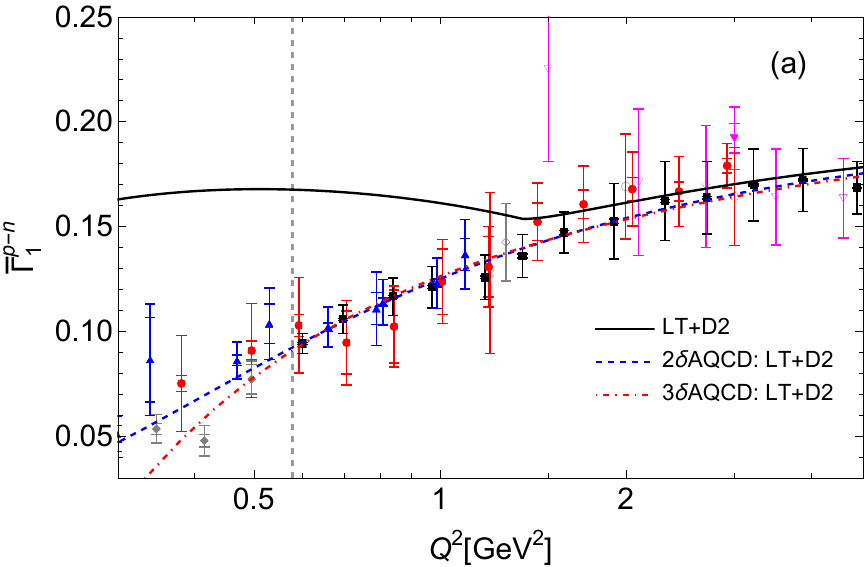}
\end{minipage}
\begin{minipage}[b]{.49\linewidth}
  \includegraphics[width=80mm,height=50mm]{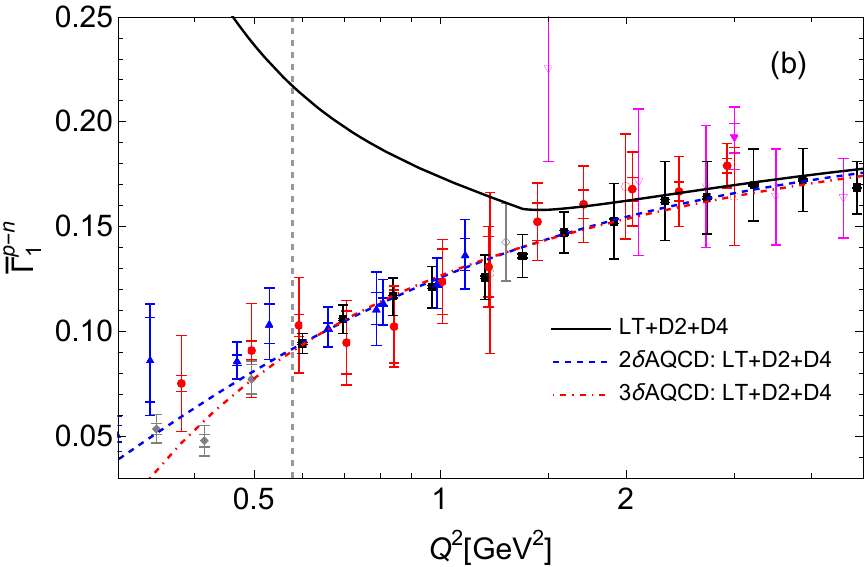}
\end{minipage} \vspace{4pt}
\caption{\footnotesize The fitted curves for the BSR ${\overline \Gamma}_1^{\rm p-n}(Q^2)$: (a) for the one-parameter fits; (b) for the two-parameter fits. The couplings used are 2$\delta$AQCD and 3$\delta$AQCD. Experimental data, and the corresponding pQCD fit (here as solid line; $Q^2_{\rm min}=1.71 \ {\rm GeV}^2$), are included for comparison. See the text for details. The vertical line indicates the value of the left end of the fit interval for the AQCD curves, $Q^2_{\rm min}=0.592 \ {\rm GeV}^2$. See Figs.~\ref{Figstasys} for identification of the different experiments each point comes from.}
\label{FigfitAQCD}
\end{figure}
In Figs.~\ref{FigfitAQCD}(a), (b), we present the resulting fitting curves ${\overline \Gamma}_1^{\rm p-n}(Q^2; {\bar f}_2)$ and ${\overline \Gamma}_1^{\rm p-n}(Q^2; {\bar f}_2, \mu_6)$ for 2$\delta$AQCD and 3$\delta$AQCD, using the corresponding central values of parameters of Eqs.~(\ref{resLTD242d})-(\ref{bf2D23d}). The corresponding central values for the other parameters were used (scheme, $\alpha_s^{\MSbar}(M_Z^2)$, $M_1$, $d_4$).
For comparison, the experimental data, and the corresponding pQCD fit (with $Q^2_{\rm min}=1.71 \ {\rm GeV}^2$), are included in the Figures. 
We recall that the fits were conducted in the $Q^2$-intervals with $Q^2_{\rm max}=4.739 \ {\rm GeV}^2$, so that $N_f=3$ QCD approaches were used throughout. Comparison of Figs.~\ref{FigfitAQCD}(a) and \ref{FigfitAQCD}(b) indicates that the 3$\delta$AQCD curves for the one-parameter fit and the two-parameter fit are very close to each other and almost cannot be distinguished from each other by eyesight. The same is true for the 2$\delta$AQCD curves for the one-parameter and the two-parameter fit. This indicates that the one-parameter fits (i.e., using OPE with LT+D2) are entirely sufficient in AQCD. In this context, we also note that the multi-$k$ approach, explained in Appendix \ref{app:ks}, keeps the extracted values of ${\bar f}_2$ in the one-parameter fits almost unchanged in comparison to the common-$k$ fit approach applied here, as can be seen from the ('$k$')-uncertainties for these quantities in Eqs.~(\ref{bf2D22d}) and (\ref{bf2D23d}).

\section{Discussion of the results, and conclusions}
\label{sec:concl}

In this work, we performed various analyses of the inelastic Bjorken polarised sum rule (BSR) ${\overline \Gamma}_1^{{\rm p-n}}(Q^2)$. The theoretical basis was the OPE (\ref{BSROPE}) truncated at the dimension $D=4$ or $D=2$ term. The canonical leading-twist QCD contribution $d(Q^2)$ was evaluated by a renormalon-based resummation, Eq.~(\ref{dres2c}), using two types of the holomorphic (AQCD) running couplings, $\alpha_s(Q^{'2})/\pi \mapsto \A(Q^{'2})$, that are free of Landau singularities: 2$\delta$AQCD and 3$\delta$AQCD couplings $\A(Q^{'2})$. This work can be regarded as a continuation of our previous work \cite{pPLB} where the analyses were performed using the same kind of resummation but with perturbative QCD (pQCD) couplings  $\alpha_s(Q^{'2})/\pi \equiv a(Q^{'2})$ that do have Landau singularities and thus a regularisation was needed, cf.~Eq.~(\ref{dres2b}). These resummations, Eqs.~(\ref{dres2b})-(\ref{dres2c}), are completely invariant under the renormalisation scale variation. The obtained theoretical (truncated OPE) expression Eq.~(\ref{BSROPE}) was then fitted to the available data points for BSR.

In general, the experimental data have too high uncertainties for the extraction of the preferred value of $\alpha_s^{\MSbar}(M_Z^2)$, especially when we include the $D=4$ term in the OPE.
It turns out that in the 2$\delta$AQCD and 3$\delta$AQCD approaches and with OPE truncated at $D=2$, the minimal $k$ (and thus the most precise fit) does exist under the variation of $\alpha_s^{\MSbar}(M_Z^2)$, and is obtained at $\alpha_s^{\MSbar}(M_Z^2) \approx 0.117$ for the central values of other ``input''parameters ($M_1$, $d_4$, etc.) and $Q^2_{\rm min}=0.592 \ {\rm GeV}^2$. This would imply that the preferred value of $\alpha_s$ is $\alpha_s^{\MSbar}(M_Z^2) \approx 0.117$. However, the minimum for $k$ is very shallow, and it disappears when $Q^2_{\rm min}$ increases, e.g., to $Q^2_{\rm min}=1 \ {\rm GeV}^2$. In the approach where both $D=2$ and $D=4$ terms are included in the OPE, we do not even obtain any minimal $k$ for any $Q^2_{\rm min}$ under the variation of $\alpha_s$. We conclude that the preferred value of $\alpha_s^{\MSbar}(M_Z^2)$ cannot be determined unambiguously, principally because of large uncertainties of the measured BSR data, especially at larger $Q^2 > 1 {\rm GeV}^2$.

Therefore, we fixed this value to $\alpha_s^{\MSbar}(M_Z^2)=0.1179$, the central world average value \cite{PDG2023}.
The resulting extracted values of the OPE fit parameters ${\bar f}_2$ and $\mu_6$ are given for the 2$\delta$AQCD case in Eqs.~(\ref{resLTD242d})-(\ref{bf2D22d}), and for the 3$\delta$AQCD case in Eqs.~(\ref{resLTD243d})-(\ref{bf2D23d}), and can be compared with values for the pQCD case obtained in Ref.~\cite{pPLB} [cf.~Eqs.~(\ref{resLTD24})-(\ref{bf2D2}) here]. The various experimental (and related) uncertainties of the extracted values, Eqs.~(\ref{resLTD242d})-(\ref{bf2D23d}), are represented by the last four terms there: ('$Q^2_{\rm min}$'), (exp.u.), (exp.c.) and ('$k$'). We see that the experimental uncertainties ('$Q^2_{\rm min}$'), (exp.u.) and (exp.c.) are in general the dominant ones, especially the correlated uncertainty (exp.c.).\footnote{A different determination of the $k$-parameter may influence this conclusion. Nonetheless, in Table \ref{tabmultik} in Appendix \ref{app:ks}, we can see that the most dominant group of experiments is that of Hall B of Jefferson Lab, and that the values of $k$ in those cases are very close to those of the common-$k$ approach (the last line in Table \ref{tabmultik}). Therefore, such a multi-$k$ approach would not significantly change the (exp.u.) and (exp.c.) uncertainties of the extracted results Eqs.~(\ref{resLTD242d})-(\ref{bf2D23d}), especially for the one-parameter fits. Nonetheless, the multi-$k$ approach in general tends to increase (exp.u.) and decrease (exp.c.).}

The various theoretical uncertainties are in general smaller than the experimental ones, sometimes with the exception of the ('$M_1$') uncertainty in AQCD cases, i.e., the uncertainty due to the variation of the spectral function of the holomorphic coupling at low scales $M_1$. Furthermore, when compared to the pQCD case Eqs.~(\ref{resLTD24})-(\ref{bf2D2}), we see that the use of AQCD couplings in general reduces the experimental uncertainties of the extracted values by large factors in comparison to the pQCD case, in some cases by one order of magnitude. Additionally, the use of AQCD couplings in the two-parameter fit gives very suppressed values of the $D=4$ parameter $\mu_6$, thus largely reducing the need to include the $D=4$ OPE term.

As mentioned in the Introduction, in our view the numerical values (and thus the operational significance) of the higher-twist quantities such as ${\bar f}_2$ and $\mu_6$ in OPE is significantly dependent on how the resummation of the leading-twist ($D=0$) part $d(Q^2)$ is performed. For example, in the Borel resummation [or any other resummation, such as Eq.~(\ref{dres2b})] of $d(Q^2)$ in pQCD, there are always renormalon ambiguities involved ($\sim 1/(Q^2)^p$) and they are reflected by the corresponding ambiguities of the higher-twist terms in the OPE. Although these ambiguities are usually fixed by a convention (e.g., by PV or related prescriptions), they still exist in principle. In our approaches of resummation of $d(Q^2)$ that involve (integrals of) IR-safe couplings $\A(t Q^2)$, presented here, an analogous kind of ambiguity appears, which is associated with the (not unique) ways of regularising the $\A$-coupling spectral function $\rho_{\A}(\sigma) = {\rm In} \A(-\sigma - i \epsilon)$ in the IR-regime ($0 < \sigma < 1 \ {\rm GeV}^2$). In our approaches, one of such ambiguity parameters can be considered to be the threshold scale $M_1$ of $\rho_{\A}(\sigma)$, cf.~Eq.~(\ref{rhoAnd}).

The (resummed) AQCD results have yet another attractive feature when compared to the (resummed) pQCD results: the preferred fit interval is considerably wider, $0.592 \ {\rm GeV}^2 \leq Q^2 \leq 4.739 \ {\rm GeV}^2$, than in the pQCD case where it was restricted to  $1.710 \ {\rm GeV}^2 \leq Q^2 \leq 4.739 \ {\rm GeV}^2$. One reason for this is that the pQCD running coupling $a(Q^{'2})$ has Landau cut singularities at low positive values $Q^{'2} \lesssim  1 \ {\rm GeV}^2$, while the AQCD couplings $\A(Q^{'2})$ are free of such singularities.

If we apply in the pQCD approach, instead of the resummation (\ref{dres2b}) in the canonical QCD part $d(Q^2)$, a simple truncated power series in $a(\mu^2)$ (TPS), the results further change significantly. For example, the TPS approach in $\MSbar$ scheme and with $\alpha_s^{\MSbar}(M_Z^2)=0.1179$, leads to significant renormalisation scale dependence, and the truncation index ($N_{\rm tr}$) dependence.\footnote{The TPS is truncated at the power $a(\mu^2)^{N_{\rm tr}}$.} If we choose $\mu^2=Q^2$, the two-parameter fit results are strongly $N_{\rm tr}$-dependent, at $N_{\rm tr}=8$ we obtain small $\chi^2=0.891$ but the values  ${\bar f}_2 \approx -0.44$ and $\mu_6 =0.43 \ {\rm GeV}^4$ are large and lead to significant cancellation effects between $D=2$ and $D=4$ BSR terms in the range $2 \ {\rm GeV}^2 < Q^2 < 3 \ {\rm GeV}^2$. On the other hand, for $N_{\rm tr} \geq 10$ we obtain very large $\chi^2 > 4$. The one-parameter fit results give for all $N_{\rm tr} \geq 3$ the values $\chi^2 >1$, and the values of $\chi^2$ increase when $N_{\rm tr}$ increases. These results strongly suggest that the TPS approach in pQCD is less reliable than the (renormalon-based) pQCD resummation approach Eq.~(\ref{dres2b}).

On the other hand, in the holomorphic QCD (i.e., AQCD) the Pad\'e-related approximants ${\cal G}_d^{[M/M]}(Q^2)$, which are constructed only from the first $2 M$ expansion coefficients $\td_n(\kappa)$ of $d(Q^2)$, are completely independent of the renormalisation scale parameter $\kappa$ and converge rapidly to the fully resummed values of $d(Q^2)$ of Eq.~(\ref{dres2c}) for all $Q^2$ when $M$ increases. These approximants in general do not work well when pQCD coupling is used, due to the Landau singularities of such a coupling.

Our results suggest that for an improved theoretical description of low-energy spacelike observables such as BSR it is important: (a) to perform the fit with resummation of $d(Q^2)$ according to Eqs.~(\ref{dres2b}) or (\ref{dres2c}) instead of using truncated expressions; (b) to use, in the resummation, for the running coupling, instead of the pQCD coupling $a(Q^{'2})$, a holomorphic coupling $\A(Q^{'2})$, i.e., a coupling that is practically equal to the (underlying) pQCD coupling $a(Q^{'2})$ at high scales $Q^{'2} > 1 \ {\rm GeV}^2$ and is regulated in the low-scale regime $Q^{'2} \lesssim 1 \ {\rm GeV}^2$ such that it has no Landau singularities there. Our analysis suggests that it would be beneficial to have the experimental uncertainties of the BSR data significantly reduced in the range $Q^2 > 0.6 \ {\rm GeV}^2$, because this would lead, in our analysis, to reduced uncertainties of the extracted parameters ${\bar f}_2$ and $\mu_6$, in various versions of AQCD as well as in pQCD. Further, this reduction would eventually lead us to be able to extract values of the coupling $\alpha_s^{\MSbar}(M_Z^2)$. Furthermore, additional data at different values of $Q^2$ are forthcoming, such as those from Jefferson Lab at 12 GeV \cite{jlab12gev} and the Electron Ion Collider \cite{eicref}.

In our work we did not consider models for inelastic BSR ${\overline \Gamma}_1^{{\rm p-n}}(Q^2)$ at very low $Q^2 < 1 \ {\rm GeV}^2$, e.g., expansions \cite{JeffL2} motivated on chiral perturbation theory or the light-front holographic QCD (LFH) \cite{LFH1,LFHBSR,LFHBSRext}. In one of our previous works \cite{ACKS} we included such low-$Q^2$ models in the analysis.\footnote{The QCD approaches applied in \cite{ACKS} at $Q^2 \gtrsim 1 \ {\rm GeV}^2$ for evaluation of $d(Q^2)$ of BSR were truncated (i.e., not resummed) series, either in pQCD or in AQCD variants.} In the present work, the main emphasis was to construct a renormalon-based extension (to all orders) of the expansion of the canonical BSR part $d(Q^2)$, and to resum it with the approach of characteristic function in the framework of AQCD variants, Eq.~(\ref{dres2c}). With these results, the corresponding OPE was fitted to the experimental data.

We performed numerical analyses (fits) with mathematica software. The mathematica programs that were constructed and used in the numerical analyses in this work are available on the web page \cite{www}, and they include the experimental data.

\begin{acknowledgments}
This work was supported in part by FONDECYT (Chile) Grants No.~1200189 and 1240329 (C.A.) and No.~1220095 (G.C.). We thank G.~Bali for a useful clarification.
\end{acknowledgments}

\appendix

\section{2$\delta$AQCD and 3$\delta$AQCD}
\label{app:AQCD}

Here we summarise the construction of 2$\delta$AQCD \cite{2dAQCD,2dAQCDb}\footnote{In \cite{2dAQCD,2dAQCDb}, we constructed 2$\delta$AQCD in a class of renormalisation schemes where only $c_2$ scheme parameter is adjustable (``3-loop'' adjustable), while here we present 2$\delta$AQCD in the P44-class of renormalisation schemes, Eqs.~(32)-(35) of \cite{pPLB}, which have $c_2$ and $c_3$ scheme parameters adjustable (``4-loop'' adjustable).}
and 3$\delta$AQCD \cite{3dAQCD,3dAQCDb}, i.e., versions $\A(Q^2)$ of the QCD coupling that have no Landau singularities and practically coincide at high $Q^2 > 1 \ {\rm GeV}^2$ with the underlying pQCD coupling $a(Q^2)$ running in a P44-renormalisation scheme, Eqs.~Eqs.~(32)-(35) of \cite{pPLB}. At low $Q^2 \lesssim 1 \ {\rm GeV}^2$, the coupling $\A(Q^2)$ is required to fulfill certain additional conditions.

The starting point is the pQCD coupling $a(Q^2)$ in a certain renormalisation scheme, for convenience the P44-scheme with chosen values of the scheme parameters $c_2$ and $c_3$, cf.~Eqs.~Eqs.~(32)-(35) of \cite{pPLB}. The resulting underlying ($N_f=3$) pQCD coupling $a(Q^2)$ is given in terms of the Lambert function, cf.~Eq.~(34) of \cite{pPLB}, it is given in terms of the Lambert function $W_{\pm 1}(z)$ (which is very convenient in practical evaluations), and it is an explicit function of any complex $Q^2$. This coupling $a(Q^2)$ has discontinuity (cut) along the real axis; this discontinuity is usually called the spectral function of the coupling
\be
\rho^{\rm (pt)}(\sigma) \equiv {\rm Im} \; a(-\sigma-i \epsilon),
\label{rhopt} \ee
which is thus again written in terms of the Lambert function $W_{\pm 1}(z)$ (and thus easily evaluated in practice). Since the coupling has Landau singularities, the spectral function is nonzero not just for $\sigma>0$ (i.e., $Q^2 = - \sigma < 0$), but also at some lower negative values $-\Lambda^2 \leq \sigma \leq 0$) (i.e., $0 \leq Q^2 = -\sigma \leq \Lambda^2$) where usually $\Lambda^2 \sim 0.1 \ {\rm GeV}^2$.

The holomorphic coupling $\A(Q^2)$ is then required to have the spectral function $\rho_{\A}(\sigma) \equiv {\rm Im} \A(-\sigma - i \epsilon)$ which coincides with the above spectral function $\rho^{\rm (pt)}(\sigma)$ for sufficiently large $\sigma > M_0^2$ ($>0$); for $0 < \sigma < M_0^2$ deviations from $\rho^{\rm (pt)}(\sigma)$ are expected; and for $\sigma <0$ we require that $\rho_{\A}(\sigma)=0$ (i.e., that there are no Landau singularities). In the range $0 < \sigma < M_0^2$ where $\rho_{\A}(\sigma)$ is not known, we parametrise it by a linear combination of $n$ Dirac delta functions [which corresponds to near-diagonal Pad\'e expression contribution in $\A(Q^2)$, see later]
\be
\rho_{\A}^{(n \delta)}(\sigma) =  \pi \sum_{j=1}^{n} {\cal F}_j \; \delta(\sigma - M_j^2)  + \Theta(\sigma - M_0^2) \rho^{\rm (pt)}(\sigma) \ .
\label{rhoAnd}
\ee
By notational convention, we have ($0 <$) $M_1^2 < M_2^2 < \cdots < M_n^2 < M_0^2$, and $M_1^2=M^2_{\rm thr}$ is interpreted as the threshold scale of the spectral function $\rho_{\A}$; it is expected to be in the range of the lowest hadronic scales, i.e., $M_1^2 \sim m_{\pi}^2$ ($\sim 10^{-2} \ {\rm GeV}^2$). On the other hand, $M_0^2$ ($\sim 1 \ {\rm GeV}^2$) can be interpreted as the pQCD-onset scale. Using the Cauchy theorem, we then obtain from the spectral function (\ref{rhoAnd}) the running coupling $\A(Q^2)$
\bea
\A^{(n \delta)}(Q^2) \left( \equiv \frac{1}{\pi} \int_0^{\infty} d \sigma \frac{\rho_{\A}(\sigma)}{(\sigma + Q^2)} \right) & = &  \sum_{j=1}^n \frac{{\cal F}_j}{(Q^2 + M_j^2)} + \frac{1}{\pi} \int_{M_0^2}^{\infty} d \sigma \frac{ \rho_1^{\rm (pt)}(\sigma) }{(Q^2 + \sigma)} \ .
\label{AQ2}
\eea
The obtained coupling has altogether $(2 n +1)$ parameters: ${\cal F}_j$ and $M_j^2$ ($j=1, \ldots, n$) and $M_0^2$.\footnote{We note that we have the Lambert scale $\Lambda_L$ in the underlying pQCD coupling, Eq.~(35) of \cite{pPLB}, and thus in $\rho^{\rm (pt)}(\sigma)$, but this scale is fixed by the chosen value of $\alpha_s^{\MSbar}(M_Z^2)$, as explained in Sec.~V of \cite{pPLB}.} They are then fixed by various conditions. The condition that the coupling should practically coincide with the underlying pQCD coupling $a(Q^2)$ at sufficiently high $|Q^2| > 1 \ {\rm GeV}^2$ is implemented in our approach in the following specific way:
\be
\A(Q^2) - a(Q^2) \sim \left( \frac{\Lambda_L^2}{Q^2} \right)^5  \qquad
(|Q^2|>\Lambda_L^2).
\label{diffAaN5}
\ee
In general, the above difference would\footnote{The difference (\ref{diffAaN5}) is $\sim (\Lambda_L^2/Q^2)^1$ in, e.g., Minimal Analytic framework (MA; named also (F)APT) \cite{Shirkov,SMS,KarSt,FAPT,ShirRev,BakRev,StRev}, i.e., the QCD variant in which the spectral function of the coupling is [instead of Eq.~(\ref{rhoAnd})]: $\rho_{\A}^{\rm (MA)}(\sigma) = \Theta(\sigma) \rho^{\rm (pt)}(\sigma)$.}
be $\sim (\Lambda_L^2/Q^2)^1$; therefore, the condition (\ref{diffAaN5}) represents four conditions.

In the 2$\delta$AQCD, we have $n=2$ and thus five parameters. We can choose a value of the threshold scale $M_1^2$ as an input, and then the four remaining parameters are fixed by the conditions (\ref{diffAaN5}). We note that in the 2$\delta$AQCD, the underlying pQCD coupling can be in any chosen P44-scheme (i.e., with any chosen values of $c_2$ and $c_3$).

In 3$\delta$AQCD, we have $n=3$ and thus seven parameters. The conditions (\ref{diffAaN5}) represent four conditions. Two additional conditions are obtained if we require that $\A(Q^2)^{(3 \delta)}$ behaves at low $Q^2$ as a specific product of the Landau gauge gluon and ghost dressing functions whose behaviour at low positive $Q^2$ was obtained by large volume lattice calculations \cite{BIMS1,BIMS2}. For details we refer to \cite{3dAQCD,3dAQCDb}. These two conditions,\footnote{These two conditions are in the (lattice-related) $N_f=3$ MiniMOM scheme (MM) \cite{MM1,MM2a,MM2b,MM3} ($c_2^{\rm MM}=9.29703$; $c_3^{\rm MM}=71.4538$) with $\MSbar$ scaling convention (LMM, i.e., Lambert MiniMOM).}
are that $\A(Q^2)$ at positive $Q^2$ achieves the local maximum at $Q^2 \approx 0.135 \ {\rm GeV}^2$ and that it behaves as $\A(Q^2) \sim Q^2$ ($\to 0$) at very low $Q^2$ ($0 < Q^2 \lesssim  0.1 \ {\rm GeV}^2$). This gives us additional two conditions, adding up to altogether six conditions. The seventh condition, necessary to fix all the seven parameters, is again the choice of the threshold scale $M_1^2$ ($\sim m_{\pi}^2$).

In Tables \ref{tab2d} and \ref{tab3d} we present the values of the parameters of the 2$\delta$AQCD and 3$\delta$AQCD coupling for the relevant cases used in this work.\footnote{We note in Table \ref{tab2d} that, when $c_2$ changes and $c_3$ is kept fixed in P44 scheme, $\Lambda_L$ changes very little, $\delta \Lambda_L \lesssim 10^{-6}$.}
\begin{table}
  \caption{Values of the parameters of the 2$\delta$AQCD coupling used in this work; the first entry is for the central case: P44-scheme with $c_2=9.$ and $c_3=20.$; $M_1=0.150$ GeV; $\alpha_s^{\MSbar}(M_Z^2)=0.1179$; the other entries are for the case when one of these parameters changes. The dimensionless parameters are $s_j=M_j^2/\Lambda_L^2$ and $f_j={\cal F}_j/\Lambda_L^2$. $\Lambda_L$ is the Lambert scale of the underlying pQCD coupling, fixed by the condition $\alpha_s^{\MSbar}(M_Z^2)=0.1179$.}
\label{tab2d}
\begin{ruledtabular}
\centering
\begin{tabular}{r|llllll}
 {\rm case} & $s_1$ & $s_2$ & $f_1$ & $f_2$ & $s_0$ & $\Lambda_{{\rm L}}$ [GeV] 
\\
\hline
{\rm central} & 0.47584 & 68.8281 & 1.71541 & 0.94206 & 95.8788 & 0.21745 \\
\hline
$c_2=11.$, $c_3=20.$ & 0.47581 & 80.6865 & 1.93086 & 1.07273 & 112.47 & 0.21745
\\
$c_2=7.6$, $c_3=20.$ & 0.47583 & 61.3845 & 1.57632 & 0.85810 & 85.4648 & 0.21745 \\
$c_2=9.$, $c_3=35.$ & 0.77003 & 121.627 & 2.92848 & 1.62931 & 169.534 & 0.17094 \\
$c_2=9.$, $c_3=5.$ & 0.22581 & 28.9299 & 0.75915 & 0.40791 & 40.2527 & 0.31566 \\
\hline
$M_1=0.250$ GeV & 1.32175 & 74.2430 & 1.77216 & 0.98111 & 103.039 & 0.21745 \\
$M_1=0.100$ GeV & 0.21148 & 67.1287 & 1.69729 & 0.92969 & 93.6321 & 0.21745 \\
\hline
$\alpha_s^{\MSbar}(M_Z^2)=0.1188$ & 0.43937 & 68.5939 & 1.71292 & 0.94036 & 95.5691 & 0.22630 \\
$\alpha_s^{\MSbar}(M_Z^2)=0.1170$ & 0.51606 & 69.0864 & 1.7181 & 0.94393 & 96.2202 & 0.20881 \\
\hline
\end{tabular}
\end{ruledtabular}
\end{table}
\begin{table}
  \caption{Values of the parameters of the 3$\delta$AQCD coupling used in this work; the first entry is for the central case: P44 LMM scheme (i.e., $c_2=9.29703$ and $c_3=71.4538$); $M_1=0.150$ GeV; $\alpha_s^{\MSbar}(M_Z^2)=0.1179$; the other entries are for the case when $M_1$ changes, or  $\alpha_s^{\MSbar}(M_Z^2)$ changes. The dimensionless parameters are $s_j=M_j^2/\Lambda_L^2$ and $f_j={\cal F}_j/\Lambda_L^2$. $\Lambda_L$ is the Lambert scale of the underlying pQCD coupling.}
\label{tab3d}
\begin{ruledtabular}
\centering
\begin{tabular}{r|llllllll}
 {\rm case} & $s_1$ & $s_2$ & $s_3$ & $f_1$ & $f_2$ & $f_3$ & $s_0$ & $\Lambda_{{\rm L}}$ [GeV] 
\\
\hline
{\rm central} & 1.79371 & 42.9853 & 607.164 & -0.586232 & 10.6669 & 6.06743 &
  827.469 & 0.11200 \\
\hline
$M_1=0.250$ GeV & 4.98252 & 16.7070 & 470.372 & -4.37049 & 13.2756 & 5.23222 &
647.009 & 0.11200 \\
$M_1=0.100$ GeV & 0.79720 & 88.0403 & 862.126 & -0.16976 & 12.2415 & 7.52131 &
1163.66 & 0.11200 \\
\hline
$\alpha_s^{\MSbar}(M_Z^2)=0.1188$ & 1.65625 & 40.3300 & 589.951 & -0.562248 & 10.5011 & 5.9629 & 804.698 & 0.11655 \\
$\alpha_s^{\MSbar}(M_Z^2)=0.1170$  & 1.94533 & 45.8556 & 625.783 & -0.61228 & 10.8448 & 6.17983 & 852.102 & 0.10755 \\
\hline
\end{tabular}
\end{ruledtabular}
\end{table}

\begin{figure}[htb] 
\centering\includegraphics[width=80mm]{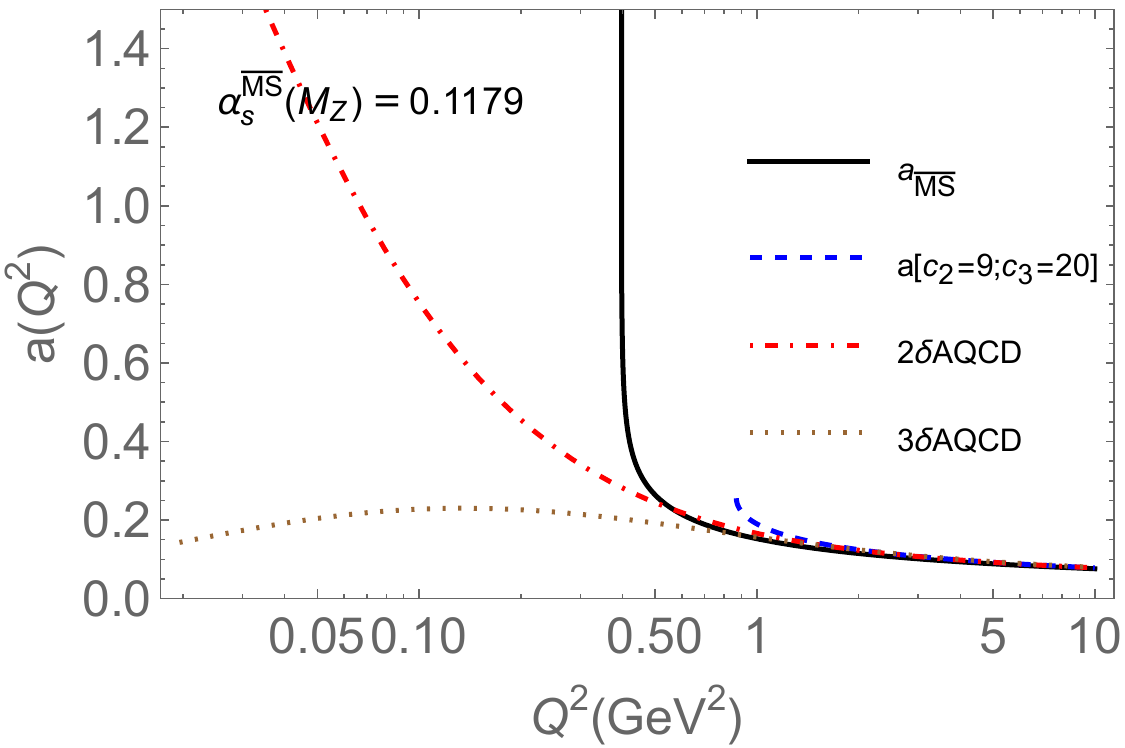}
\caption{\footnotesize Various ($N_f=3$) running couplings: pQCD coupling  $a(Q^2)_{\MSbar}$ in the 5-loop $\MSbar$ scheme; 2$\delta$AQCD coupling $\A(Q^2)^{(2 \delta)}$ and its underlying pQCD coupling $a(Q^2)$, both in the P44 scheme with $c_2=9.$ and $c_3=20.$; 3$\delta$AQCD coupling $\A(Q^2)^{(3 \delta)}$, in the (lattice-related) P44 LMM scheme.}
\label{FigA}
\end{figure}

In Fig.~\ref{FigA} we present the behaviour of various ($N_f=3$) running couplings: pQCD coupling $a(Q^2)_{\MSbar}$ in the 5-loop $\MSbar$ scheme; pQCD coupling $a(Q^2)$ in the P44-scheme with $c_2=9.$ and $c_3=20.$; 2$\delta$AQCD coupling $\A(Q^2)^{(2 \delta)}$ (in the mentioned P44 scheme with $c_2=9.$ and $c_3=20.$); and 3$\delta$AQCD coupling  $\A(Q^2)^{(3 \delta)}$ in the P44 LMM scheme ($c_2=9.29703$ and $c_3=71.4538$).
The coupling $a(Q^2)$ has Landau cut for $Q^2 < 0.87 \ {\rm GeV}^2$, and $a(Q^2)_{\MSbar}$ for $Q^2 < 0.40 \ {\rm GeV}^2$. We note that at the corresponding branching points  $a(Q^2)$ is finite and $a(Q^2)_{\MSbar}$ is infinite. The AQCD couplings have neither Landau cuts nor infinities. The coupling $\A(Q^2)^{(2 \delta)}$ grows at decreasing $Q^2$ and reaches a large, but finite value $\A(0)^{(2 \delta)}=3.6970$ at $Q^2=0$. All couplings correspond to the reference value $\alpha_s^{\MSbar}(M_Z^2;N_f=5)=0.1179$, and the AQCD couplings have the spectral threshold scale $M_1=0.150$ GeV.

\begin{figure}[htb] 
\begin{minipage}[b]{.49\linewidth}
\includegraphics[width=80mm,height=50mm]{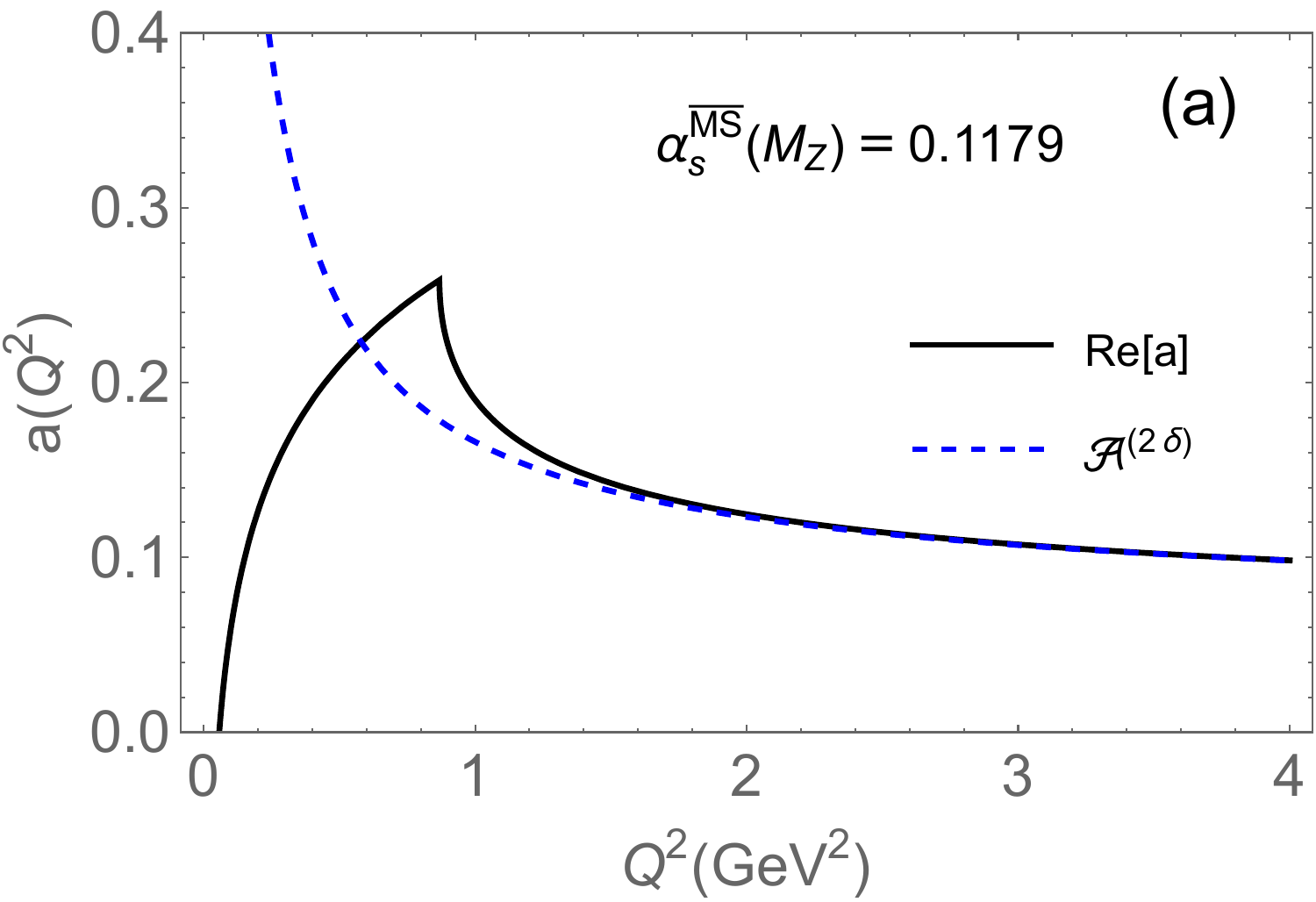}
\end{minipage}
\begin{minipage}[b]{.49\linewidth}
\includegraphics[width=80mm,height=50mm]{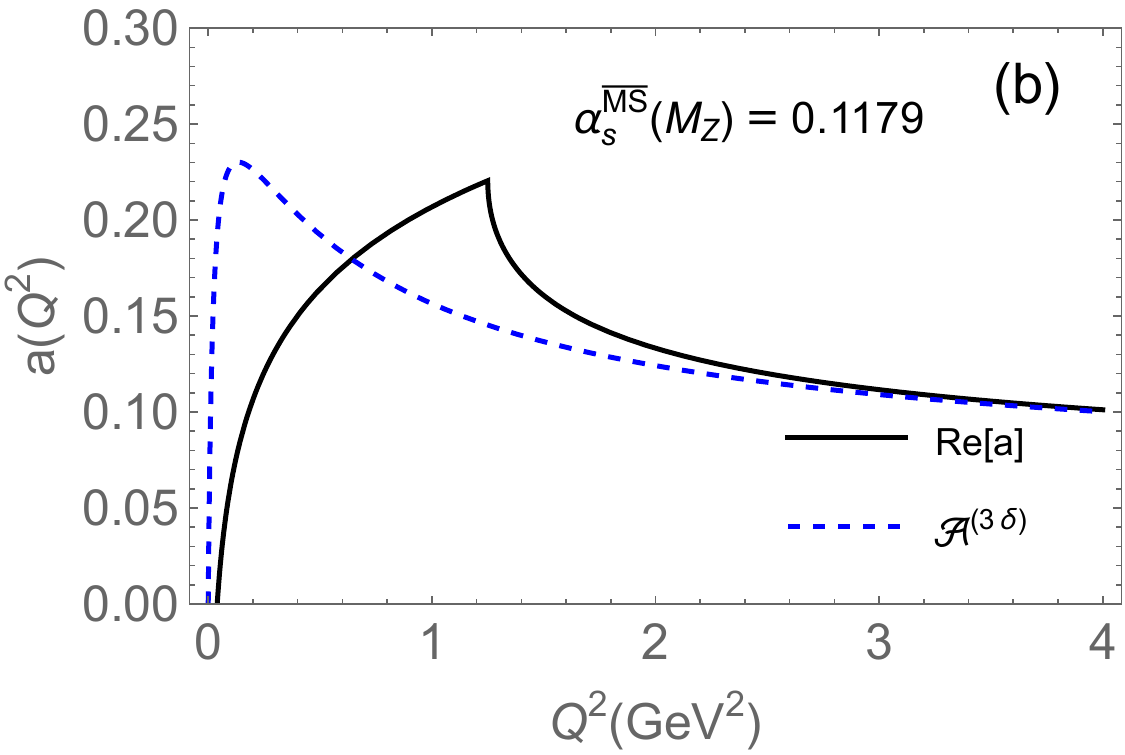}
\end{minipage} \vspace{4pt}
\caption{\footnotesize The AQCD running coupling $\A(Q^2)$ and their corresponding underlying pQCD coupling $a(Q^2)$: (a) for 2$\delta$AQCD (in the scheme P44 with $c_2=9.$ and $c_3=20.$); (b) for 3$\delta$AQCD (in the P44 LMM scheme). The coupling $a(Q^2)$ has Landau cut at $0 \leq Q^2 <0.87 \ {\rm GeV}^2$ in the 2$\delta$ case, and at $0 \leq Q^2 <1.25 \ {\rm GeV}^2$ in the 3$\delta$ case. Both these couplings $a(Q^2)$ become complex on the Landau cut. Therefore, real part ${\rm Re} \ a(Q^2)$ was presented in those intervals.}
\label{FigA2d3d}
\end{figure}

Further, in Figs.~\ref{FigA2d3d}(a),(b), we present separately the 2$\delta$AQCD and 3$\delta$AQCD running couplings; for comparison, we include the corresponding underlying pQCD couplings $a(Q^2)$.

The expansions in AQCD are made in a completely analogous way as in pQCD, Eq.~(\ref{dlpt}), i.e., in terms of the logarithmic derivatives $\tA_n$ that are the AQCD analogues of the $\ta_n$ logarithmic derivatives of pQCD coupling [cf.~Eq.~(\ref{tan})]
\be
{\tA}_{n}(\mu^2) \equiv \frac{(-1)^{n-1}}{(n-1)! \beta_0^{n-1}} \left( \frac{d}{d \ln \mu^2} \right)^{n-1} \A(\mu^2) \qquad (n=1,2,\ldots).
\label{tAn} \ee
\be
d(Q^2) = \A( \kappa Q^2) + {\td}_1(\kappa) \; {{\tA}}_2(\kappa Q^2) + \ldots + {\td}_n(\kappa) \; {{\tA}}_{n+1}(\kappa Q^2) + \ldots,
\label{dlApt} \ee
This expansion in AQCD was introduced in \cite{CV}, and it has the form completely analogous to the pQCD expansion in logarithmic derivatives $\ta_n$ Eq.~(\ref{dlpt}).\footnote{\label{tAnu}The extension of the logarithmic derivative, $\tA_{\nu}(Q^2)$, for noninteger $\nu$ in any AQCD was constructed in \cite{GCAK}, as well as the coupling $\A_{\nu}(Q^2)$ that is the AQCD analogue of the power $a(Q^2)^{\nu}$ (for $\nu$ noninteger, $-1 < \nu$). For the Minimal Analytic (MA) QCD, the extended logarithmic derivatives $\tA_{\nu}^{\rm (MA)}(Q^2)$ were constructed as explicit functions at one-loop order in \cite{FAPT} and at any loop order in \cite{Kotikov}.}$^,$\footnote{For some applications of various AQCD to QCD phenomenology, see, e.g., \cite{ShirkEPJC,Nest1,Nest2,NestBook,GCAK,ACKS,KotBSR,CAGCUps,CASM,Mirj,Nestamu,GCRKamu}.}
Another, more efficient, sequence of approximants are specific type of approximants, ${\cal G}^{[M/M]}_{d}(Q^2)$, which are related to the diagonal Pad\'e approximants, and are constructed only on the basis of the knowledge of the first $2 M$ coefficients of the series in $\ta_n$ in pQCD, or equivalently of the series (\ref{dlApt}) in AQCD: $\td_j(\kappa)$ ($j=0, 1, \ldots, 2 M-1$). The approximants ${\cal G}^{[M/M]}_{d}(Q^2)$ were proposed in \cite{dBGpt} in the context of pQCD, and were later applied in variants of AQCD in \cite{dBG1,dBG2,dBG3,renmod} to the (spacelike) Adler function and to related QCD observables. They are completely independent of the renormalisation scale parameter $\kappa$.\footnote{The corresponding diagonal Pad\'e approximants are $\kappa$-invariant only at the one-loop level approximation \cite{GardiPA}.} These approximants have in AQCD the form
\be
{\cal G}^{[M/M]}_{d}(Q^2)_{{\rm AQCD}} =    \sum_{k=1}^{M} \tal_j \A(\kappa_j Q^2),
\label{GMM} \ee
and fulfill the approximant precision relation
\be
d(Q^2) - {\cal G}^{[M/M]}_{d}(Q^2)_{{\rm AQCD}} = {\cal O}(\tA_{2 M +1}) = {\cal O}(a^{2 M+1}).
\label{GMMprec} \ee

In pQCD, the same expression (\ref{GMM}) is valid, but with $\A(\kappa_j Q^2)$ replaced by the pQCD coupling $a(\kappa_j Q^2)$.

We note that the approximant (\ref{GMM}) contains in total $2 M$ parameters: $\tal_j$ and $\kappa_j$, which are determined by the first $2 M$ expansion coefficients $\td_n(\kappa)$ ($n=0,1,\ldots, 2 M-1$).\footnote{
  We note that $\td_0=1$, and therefore $\sum \tal_j = 1$.}
As noted in \cite{dBGpt}, the application of these approximants in pQCD gives for spacelike QCD observables $d(Q^2)$ in general relatively unstable results, especially when $M$ increases, and this is so because some of the $\kappa_j$ coefficients are very small and then $a(\kappa_j Q^2)$ is on Landau cut singularities. This problem does not appear in QCD with holomorphic couplings (AQCD), as noted in \cite{dBG1,dBG2,dBG3,renmod}, because $\A(\kappa_j Q^2)$ coupling has no Landau singularities.

In Table \ref{tabdBG} we present the values of the parameters $\kappa_j$ and $\tal_j$ of the approximants ${\cal G}^{[M/M]}_{d}(Q^2)$ for canonical BSR $d(Q^2)$, for $M=2$ and $M=3$, for 2$\delta$QCD (in the P44 scheme with $c_2=9.$ and $c_3=20.$) and for 3$\delta$AQCD (in the P44 LMM scheme, i.e., with $c_2=9.29703$ and $c_3=71.4538$).
\begin{table}
  \caption{Values of the parameters $\tal_j$ and $\kappa_j$ of the (diagonal Pad\'e-related) approximants ${\cal G}^{[M/M]}_{d}(Q^2)$, for 2$\delta$AQCD and 3$\delta$AQCD (in their respective P44-schemes), for $M=2$ and $M=3$.}
\label{tabdBG}
\begin{ruledtabular}
\centering
\begin{tabular}{r|r|lll|lll}
 {\rm QCD variant} & $M$ & $\tal_1$ & $\tal_2$ & $\tal_3$ & $\kappa_1$ & $\kappa_2$ & $\kappa_3$ 
\\
\hline
2$\delta$AQCD & 2 & -0.011789 & 1.011789 & -- & 462.707872 & 0.222567 & -- \\
2$\delta$AQCD & 3 & -0.045853 & 0.013012 & 1.032841 & 20.911126 & 0.0033448 & 0.263115 \\
\hline
3$\delta$AQCD & 2 & -0.044854 & 1.044854 & -- & 13.47291 & 0.243514 & --  \\
3$\delta$AQCD & 3 & -0.095297 & 0.008018 & 1.087279 & 4.781674 & 0.006406 & 0.275173 \\
\hline
\end{tabular}
\end{ruledtabular}
\end{table}

\section{The experimental uncorrelated uncertainty of the extracted parameter values}
\label{app:expu}

We summarise here the formula for the uncorrelated experimental uncertainties for the two-parameter fit (${\bar f}_2$ and $\mu_6$) of the (inelastic) BSR. This is a special case of the approach given in App. of Ref.~\cite{Bo2011} (and App.~D of \cite{ACT2023}).\footnote{The latter approach is valid even in the case when we have known nonzero correlations of experimental data.}

The fit consist of minimising the expression $\chi^2$ Eq.~(\ref{chi2}), where the two OPE parameters $p_1={\bar f}_2$ and $p_2=\mu_6$ are varied.
The corresponding variations of these two parameters, due to the experimental uncertainties $\sigma(Q^2_j)$ at each point $Q^2_j$, are
\bes
\label{dpjdpj}
\bea
\langle \delta p_1 \delta p_1 \rangle & = & (A^{-1})_{1,1},
\label{dp1dp1} \\
\langle \delta p_2 \delta p_2 \rangle & = & (A^{-1})_{2,2},
\label{dp2dp2}
\eea \ees
where the $2 \times 2$ matrix $A$ is
\be
A_{\ell,m} = \sum_{j=j_{\rm min}}^{j_{\rm max}} \frac{\partial {\overline \Gamma}_1^{{\rm p-n}}(Q^2_j)}{\partial p_{\ell}} \frac{\partial {\overline \Gamma}_1^{{\rm p-n}}(Q^2_j)}{\partial p_m} \frac{1}{\sigma(Q^2_j)^2} \qquad (\ell, m=1,2).
\label{Alm} \ee
Then the (uncorrelated) experimental uncertainties of the extracted values of $p_1={\bar f}_2$ and of $p_2 = \mu_6$ are
\bes
\bea
\delta {\bar f}_2 ({\rm exp.u.}) &=& \sqrt{\langle \delta p_1 \delta p_1 \rangle}
  = \sqrt{(A^{-1})_{1,1}},
\label{delbf2} \\
\delta \mu_6 ({\rm exp.u.}) &=& \sqrt{\langle \delta p_2 \delta p_2 \rangle}
  = \sqrt{(A^{-1})_{2,2}},
  \label{delmu6}
\eea \ees
In the case of the one-parameter fit (${\bar f}_2$), analogous formulas hold, $A$ is in such a case only a number ($1 \times 1$ matrix).

\section{Multi-$k$ approach}
\label{app:ks}

In the main text, we described how the value of the (common) parameter $k$, which appears in Eqs.~(\ref{chi2})-(\ref{sig2}), was determined for each of the four fit approaches (2$\delta$QCD and 3$\delta$QCD, one- and two-parameter fit). However, since the value of $k$ is a measure of the contributions of the uncorrelated vs correlated experimental uncertainties of the BSR measured values, it would be more correct to fix the value of $k$ for each experiment separately.

Therefore, here we repeat the mentioned procedure, but for each experiment separately (i.e., multi-$k$ approach). For simplicity, we use in the fits the same value $Q^2_{\rm min}=0.592 \ {\rm GeV}^2$ (and $Q^2_{\rm max}=4.739 \ {\rm GeV}^2$) as in the common-$k$ approach of the main text. It turns out that the data in this $Q^2$-interval have: 25 points for the experiments at Hall B (Jefferson Lab); 3 points for the experiment at Hall A (Jefferson Lab); 3 points at SLAC; 5 points at HERMES (DESY); and 1 point each at both Hall C (Jefferson Lab) and COMPASS (CERN).\footnote{In the case of the Jefferson Lab data, we joined various related experiments (cf.~Figs.~\ref{Figstasys}) into common groups (experiments): Hall B has EG1a (2 points), EG1b (10 points), EG1-DVCS (13 points) (and EG4 has no points in the considered $Q^2$-interval); Hall A has E94010 (3 points) (and E97110 has no points in the considered $Q^2$-interval); Hall C has JLABRSS (1 point). The combined group E94010/EG1a results (cf.~Figs.~\ref{Figstasys}) has Deuterium target points (EG1a, Hall B) and ${^3}{\rm He}$ target points (E94010, Hall A), cf.~Table I of \cite{JeffL1}.}

It turns out that the statistical and systematic uncertainties of these data are in general the smallest for the Hall B data, where the number of points is also by far the largest. Therefore, we expect that Hall B data will have dominant influence over the extracted values of the fit parameters. In practice, this indeed turns out to be so. In Table \ref{tabmultik}, we show the obtained values of the $k$ parameter for each experiment and each fit approach. We note that whenever our approach formally gives $k <0$, we choose $k=0$ in such a case (because we must have $k \geq 0$).
\begin{table}
  \caption{Values of the parameters $k$ for each experiment and each fit-approach. The second column shows the number of points $N_{\rm pts}$ in the considered $Q^2$-interval. When $N_{\rm pts}=1$, we take $k=1$. For comparison, in the last line are the results for $k$ in the common-$k$ approach used in the main text.}
\label{tabmultik}
\begin{ruledtabular}
\centering
\begin{tabular}{r|c|ll|ll}
 {\rm experiment} & $N_{\rm pts}$ & $k (2\delta; {\rm 2p})$ & $k (2\delta; {\rm 1p})$  & $k (3\delta; {\rm 2p})$ & $k (3\delta; {\rm 1p})$
\\
\hline
Hall B  & 25 & 0.1351 & 0.1273 &  0.1648 & 0.1550 \\
Hall A &  3 & 0 & 0 & 0 & 0.0004 \\
SLAC & 3 & 0 & 0 & 0 & 0 \\
HERMES & 5 & 0 & 0 & 0 & 0 \\
\hline
Hall C & 1 & 1 & 1 & 1 & 1 \\
COMPASS & 1 & 1 & 1 & 1 & 1 \\
\hline
common-$k$ & 38 & 0.1342 & 0.1295 & 0.1657 & 0.1583 \\
\hline
\end{tabular}
\end{ruledtabular}
\end{table}
When we compare the obtained values of $k$ from the (dominant) Hall B experimental data with those obtained in the common-$k$ approach (the last line in the Table), we see that the values of $k$ change only a little. Furthermore, when performing the fit (minimisation of $\chi^2$) in the multi-$k$ approach, we obtain for the case of the central values of input parameters ($\alpha_s^{\MSbar}(M_Z^2)=0.1179$; $M_1=0.150$ GeV; etc.) the following values of the extracted parameters:
\begin{enumerate}
\item
  for 2$\delta$AQCD with two-parameter fit: ${\bar f}_2^{(2 \delta)}= -0.0084$ and $\mu_6^{(2 \delta)}=0.0000 \ {\rm GeV^4}$;
\item
  for 2$\delta$AQCD with one-parameter fit: ${\bar f}_2^{(2 \delta)}=-0.0081$;
\item  
  for 3$\delta$AQCD with two-parameter fit: ${\bar f}_2^{(3 \delta)}=-0.2689$; $\mu_6^{(3 \delta)}=+0.0019 \ {\rm GeV^4}$;
\item
  for 3$\delta$AQCD with one-parameter fit: ${\bar f}_2^{(3 \delta)}=-0.2547$.
\end{enumerate}
When we compare with the extracted central values in the common-$k$ approach in the main text, Eqs.~(\ref{resLTD242d})-(\ref{bf2D22d}) and (\ref{resLTD243d})-(\ref{bf2D23d}), we see that the difference is moderate. Especially in the one-parameter fits, the difference is very small. In the two-parameter fits, the difference is somewhat larger, but it is always significantly less than the uncorrelated experimental uncertainties (exp.u.), cf.~Eqs.~(\ref{resLTD242d}) and (\ref{resLTD243d}), which in turn are significantly less than correlated experimental uncertainties (exp.c.).

\end{document}